\documentclass[sigconf]{acmart}
\usepackage{booktabs} % For formal tables
\usepackage{balance}

\usepackage[utf8]{inputenc}
\usepackage{multirow}
\usepackage{microtype}
\usepackage{todonotes}

\usepackage[toc,acronym,nowarn]{glossaries}
\newacronym{OS}{OS}{Operating System}

\newacronym{IoT}{IoT}{Internet of Things}
\newacronym{TCB}{TCB}{Trusted Computing Base}
\newacronym{IPMI}{IPMI}{Intelligent Platform Management Interface}
\newacronym{BMC}{BMC}{Baseboard Management Controller}
\newacronym{AWDT}{AWDT}{Authenticated Watchdog Timer}
\newacronym{MPU}{MPU}{Memory Protection Unit}
\newacronym{DICE}{DICE}{Device Identifier Composition Engine}
\newacronym{DoS}{DoS}{Denial of Service}
\newacronym{DDoS}{DDoS}{Distributed Denial of Service}
\newacronym{TCG}{TCG}{Trusted Computing Group}
\newacronym{COTS}{COTS}{Commercial Off-The-Shelf}
\newacronym{MCU}{MCU}{Microcontroller Unit}
\newacronym{FMC}{FMC}{Flash Memory Controller}
\newacronym{HRNG}{HRNG}{Hardware Random Number Generator}
\newacronym{TRNG}{TRNG}{True Random Number Generator}
\newacronym{CDI}{CDI}{Compound Device Identifier}
\newacronym{UDS}{UDS}{Unique Device Secret}
\newacronym{PKI}{PKI}{Public Key Infrastructure}
\newacronym{KDF}{KDF}{Key Derivation Function}
\newacronym{HMAC}{HMAC}{Hash-based MAC}
\newacronym{BLE}{BLE}{Bluetooth Low Energy}
\newacronym{CSR}{CSR}{Certificate Signing Request}
\newacronym{SPI}{SPI}{Serial Peripheral Interface}
\newacronym{ISP}{ISP}{Internet Service Provider}
\newacronym{TPM}{TPM}{Trusted Platform Module}
\newacronym{TIMA}{TIMA}{TrustZone-based Integrity Measurement Architecture}
\newacronym{SRTM}{SRTM}{Static Root of Trust for Measurement}
\newacronym{DRTM}{DRTM}{Dynamic Root of Trust for Measurement}
\newacronym{TXT}{TXT}{Trusted Execution Technology}
\newacronym{SGX}{SGX}{Software Guard Extensions}
\newacronym{VM}{VM}{Virtual Machine}
\newacronym{SEV}{SEV}{Secure Encrypted Virtualization}
\newacronym{PSP}{PSP}{Platform Security Processor}
\newacronym{PUF}{PUF}{Physical Unclonable Function}
\newacronym{HV}{HV}{Hypervisor}
\newacronym{IDS}{IDS}{Intrusion Detection System}
\newacronym{TEE}{TEE}{Trusted Execution Environment}
\newacronym{WWDT}{WWDT}{Windowed WDT}
\newacronym{LoC}{LoC}{Lines of Code}
\newacronym{SCB}{SCB}{System Control Block}
\newacronym{NSC}{NSC}{Non-Secure Callable}
\makeglossaries
\glsdisablehyper % disable hyperlinks to nowhere...
\usepackage{xspace}

\newcommand{\Title}{The Lazarus Effect: Healing Compromised Devices in the Internet of Small Things}
\newcommand{\name}{Lazarus\xspace}
\newcommand{\namecore}{Lazarus Core\xspace}
\newcommand{\nametrigger}{TEETrigger\xspace}
\newcommand{\diceup}{DICE++\xspace}
\newcommand{\hub}{hub\xspace}
\newcommand{\nxp}{NXP LPC5500\xspace}
\newcommand{\nxps}{NXP LPC55S69\xspace}
\newcommand{\dicepptbl}{\textit{DICE++}\xspace}
\newcommand{\namecoretbl}{\textit{LZ Core}\xspace}
\newcommand{\patchertbl}{\textit{CP}\xspace}
\newcommand{\udtbl}{\textit{UD}\xspace}

\usepackage{graphicx}
\DeclareGraphicsExtensions{.pdf,.jpeg,.png}

\usepackage{hyperref}
\hypersetup{
  colorlinks=true, linkcolor=black, citecolor=black, urlcolor=black,
  pdftitle={\Title}
}

\usepackage[all]{hypcap}
\usepackage{fixltx2e}
\usepackage{enumitem} % fixes overfull hboxes in description envs
\usepackage{nicefrac}
\usepackage{amssymb}
\usepackage{pbox}
\usepackage{siunitx}

\makeatletter
\let\orgdescriptionlabel\descriptionlabel
\renewcommand*{\descriptionlabel}[1]{%
  \let\orglabel\label
  \let\label\@gobble
  \phantomsection
  \edef\@currentlabel{#1\unskip}%
  \let\label\orglabel
  \orgdescriptionlabel{#1}%
}
\makeatother

\begin{document}
\fancyhead{}

% acm
\copyrightyear{2020}
\acmYear{2020}
\setcopyright{acmlicensed}
\acmConference[ASIA CCS '20]{Proceedings of the 15th ACM Asia Conference on Computer and Communications Security}{October 5--9, 2020}{Taipei, Taiwan}
\acmBooktitle{Proceedings of the 15th ACM Asia Conference on Computer and Communications Security (ASIA CCS '20), October 5--9, 2020, Taipei, Taiwan}
\acmPrice{15.00}
\acmDOI{10.1145/3320269.3384723}
\acmISBN{978-1-4503-6750-9/20/10}

\title{\Title}

% \author{Anonymized for Review}

% llncs
% \author{Manuel Huber\inst{1}\orcidID{0000-0003-0829-6902} \and
%  \author{Manuel Huber \and
%  Stefan Hristozov \and
%  Simon Ott \and
%  Vasil Sarafov
%  \institute{Fraunhofer AISEC, Garching near Munich, Germany \\
%  \email{firstname.lastname@aisec.fraunhofer.de}
%  }
%

% acm
% \author{Mathias Morbitzer, Manuel Huber, Julian Horsch and Sascha Wessel}
% \affiliation{
%   \institution{Fraunhofer AISEC}
%   \city{Garching near Munich}
%   \state{Germany}
% }
% \email{{firstname.lastname}@aisec.fraunhofer.de}

%%%TODO use this
 \author{Manuel Huber}
 \orcid{0000-0003-0829-6902}
 \affiliation{
   \institution{Fraunhofer AISEC}
   \city{Garching near Munich}
   \country{Germany}
 }
 \email{mahuber@microsoft.com}

 \author{Stefan Hristozov}
 %\orcid{}
 \affiliation{
   \institution{Fraunhofer AISEC}
   \city{Garching near Munich}
   \country{Germany}
 }
 \email{stefan.hristozov@aisec.fraunhofer.de}

 \author{Simon Ott}
 %\orcid{}
 \affiliation{
   \institution{Fraunhofer AISEC}
   \city{Garching near Munich}
   \country{Germany}
 }
 \email{simon.ott@aisec.fraunhofer.de}
 
 \author{Vasil Sarafov}
 %\orcid{}
 \affiliation{
   \institution{Fraunhofer AISEC}
   \city{Garching near Munich}
   \country{Germany}
 }
 \email{vasil.sarafov@aisec.fraunhofer.de}

 \author{Marcus Peinado}
 %\orcid{}
 \affiliation{
   \institution{Microsoft Research}
   \city{Redmond}
   \country{United States}
 }
 \email{marcuspe@microsoft.com}
\begin{abstract}
We live in a time when billions of IoT devices are being
deployed and increasingly relied upon. This makes ensuring their
availability and recoverability in case of a compromise a
paramount goal. The large and rapidly growing number of deployed
IoT devices make manual recovery impractical, especially if the
devices are dispersed over a large area. Thus, there is a need
for a reliable and scalable remote recovery mechanism that works
even after attackers have taken full control over devices,
possibly misusing them or trying to render them useless.

To tackle this problem, we present \name, a system that enables
the remote recovery of compromised IoT devices. With \name, an
IoT administrator can remotely control the code running on IoT
devices unconditionally and within a guaranteed time bound. This
makes recovery possible even in case of severe corruption of the
devices' software stack. We impose only minimal hardware
requirements, making \name applicable even for low-end
constrained off-the-shelf IoT devices. We isolate \name's
minimal recovery trusted computing base from untrusted software
both in time and by using a trusted execution environment. The
temporal isolation prevents secrets from being leaked through
side-channels to untrusted software. Inside the trusted
execution environment, we place minimal functionality that
constrains untrusted software at runtime.

We implement \name on an ARM Cortex-M33-based microcontroller
in a full setup with an IoT hub, device provisioning and secure
update functionality. Our prototype can recover compromised
embedded OSs and bare-metal applications and prevents attackers
from bricking devices, for example, through flash wear out. We
show this at the example of FreeRTOS, which requires no
modifications but only a single additional task. Our evaluation
shows negligible runtime performance impact and moderate memory
requirements.

\end{abstract}
 %dominance primitive; embedded security}

%
% TODO



%avoid line breaks in mail address
%\begingroup
%\mathchardef\UrlBreakPenalty=10000
%\def\UrlFont{\small} %smaller email
\maketitle
%\endgroup

% llncs
% \begin{abstract}
% \input{abstract}
% \keywords{trusted computing; cyber resilience; recovery; dominance primitive; embedded security}
% \end{abstract}

%\newpage
\section{Introduction}

With the \gls{IoT} becoming increasingly pervasive and a focal
topic in computing, more and more \gls{IoT} devices are rolled
out. Driven by cost savings and short product development
cycles, a vast number of \gls{IoT} business use cases have
emerged, making \gls{IoT} a disruptive technology. Examples are
home automation \cite{smarthome}, farming
\cite{smart_livestock}, sensor networks
\cite{raghavendra2006wireless}, Car2X \cite{car2X}, industrial
\gls{IoT} \cite{Tiwari:2007:EWS:1210669.1210670}, or smart
devices like tools \cite{hilti}, traffic lights \cite{sierra} or
vending machines \cite{vendingMachines}, to name only a few.
These devices are typically connected via (wireless) networks to
a \hub, a back-end server located in the cloud or
managed by an enterprise. The growing importance of \gls{IoT}
deployments for public safety and business processes makes them
an attractive target for attackers. This has been demonstrated
by a large number of attacks \cite{Nawir2016, Deogirikar2017},
such as the Mirai \cite{mirai} or the Hajime botnet
\cite{hajime}.

An important property of the \gls{IoT} domain is that many
devices with identical software stacks and configurations can
likely be found in the field. This makes identified
vulnerabilities or misconfigurations highly scalable, allowing
attackers to potentially compromise a large number of devices.
Especially small and cheap devices, oftentimes poorly secured,
can be found in large numbers \cite{MCUMarket}. Even more
alarming is the threat of attackers making such devices refuse
communication and updates from the \hub or permanently damaging
them. In consequence, devices have to be manually recovered by
replacing them or resetting them with clean software. As
billions of devices (managed by a much smaller number of
administrators) are deployed (often geographically dispersed),
manual recovery becomes completely impractical.

As a solution to this problem, several cyber-resilient \gls{IoT}
architectures have been proposed \cite{Xu2019, tcgcyres,
8714822}. These architectures enable remote recovery of infected
\gls{IoT} devices within a time bound regardless of compromise.
State-of-the-art architectures \cite{Xu2019, tcgcyres, 8714822}
employ a minimal, early-boot recovery \gls{TCB} and a reset
trigger that preempts compromised software. The recovery
\gls{TCB} ensures that only software authorized by the \hub runs
on the device. If no such software is present, the recovery
\gls{TCB} downloads the software from the \hub and replaces the
existing outdated or compromised stack. The reset trigger
ensures that a reset into the recovery \gls{TCB} will eventually
happen even if software actively refuses the reset. Both the
recovery \gls{TCB} and reset trigger are isolated from untrusted
software at runtime. This prevents malware from modifying the
recovery \gls{TCB} and from interfering with the reset trigger.
An example of a reset trigger is the \gls{AWDT} \cite{Xu2019}.
After having been initialized, the \gls{AWDT} will force a
device reset after a certain time period, unless it is serviced
with cryptographically protected ``tickets''. These tickets
cannot be forged by software on the device but only be issued by
the \hub. As a consequence, as soon as the \hub stops issuing
tickets, e.g., when a zero-day vulnerability becomes known, or
when the device behaves suspiciously, the \gls{AWDT} will
eventually time out and trigger a reset. A similar timer called
``latchable WDT'' \cite{tcgcyres} will power cycle a device
within a specified time interval after its activation.

With resilient \gls{IoT} architectures, we envision a world of
self-healing \gls{IoT} deployments where all
\gls{IoT} devices can be reliably recovered even if they are
compromised. However, existing cyber-resilient \gls{IoT}
architectures which target primarily higher-end \gls{IoT}
devices
do not address several critical real-world problems
which must be solved to fulfill this vision even for low-cost
microcontrollers.
This includes

\textbf{1) Hardware requirements.} Current designs require
hardware that typically does not exist or would incur additional
cost on weaker, low-cost \gls{COTS} devices such as \glspl{MCU}.
Examples are ``storage latches'' to write-protect the recovery
\gls{TCB} from untrusted software, or the \gls{AWDT} peripheral
that was realized previously as a separate \gls{MCU} and
attached to the main board \cite{Xu2019}. For small
\gls{MCU}-based \gls{IoT} devices, this makes the \gls{AWDT} a
feature as complex as the class of devices it is intended to
protect. For storage latches, CIDER \cite{Xu2019} relied on eMMC
memory chips which are relatively complex and expensive and
typically not available on small \gls{MCU}-based devices.

\textbf{2) Preventing attackers from disabling devices.}
Existing work neglects attacks in which the adversary tries to
make a device unavailable. Examples include entering low-power
or off states in which the reset trigger becomes inactive,
wearing out flash memory \cite{Templeman2012}, or disabling the
networking hardware which is needed to replace corrupted
software on the device.

\textbf{3) Updates of the recovery \gls{TCB}.} After devices
have been deployed, vulnerabilities in the recovery \gls{TCB}
itself may be discovered, or cryptographic requirements may
change. It is essential to patch such vulnerabilities in a
timely manner before they can be exploited by attackers.

This paper presents the \emph{\name} system, a cyber-resilient
\gls{IoT} architecture that solves these three problems.
\name targets low-cost \gls{COTS} \glspl{MCU}.
We take advantage of the recent addition of a general-purpose
\gls{TEE} to low-end ARM \gls{MCU}s in the ARMv8-M
architecture.
As the ARMv8-M replaces earlier models, we expect \glspl{TEE} to
become widely available in low-end \glspl{MCU} in the coming
years.
A key observation in the design of \name is that the
security hardware required by cyber-resilient architectures can
be emulated by
software running inside a \gls{TEE}. Furthermore, the \gls{TEE}
can mediate access by untrusted software to critical system
state.

In particular, we implement storage write protection latches and
the \gls{AWDT} in software running in the \gls{TEE}.
We call our reset trigger ``\nametrigger.''
With this design, we enable the \gls{AWDT} to be realized on
existing \gls{COTS} devices and do not make assumptions on
trusted peripherals for storage latches and reset triggers.
Our design prevents attackers from making devices permanently
unavailable by interposing between untrusted software and
peripherals that are critical for availability.
We call these ``critical peripherals'' and use the \gls{TEE} to
regulate or block access by untrusted software, preventing their
misuse.
Finally, existing \gls{IoT} architectures use the \gls{DICE} for
authentication and attestation to verify whether authorized
software is present on the device.
This enables the \hub to authenticate the deployed software
stack.
However, as the hash of the boot code is part of the \gls{DICE}
attestation identity, an update of the recovery \gls{TCB} will
give the device a new \gls{DICE} identity
that the \hub cannot predict.
This makes it impossible for the \hub to recognize device
identities after a TCB update and to verify if an update was
actually installed or not.
To overcome this problem, we propose an extension to \gls{DICE},
which we call \diceup.
Our extension enables devices to provide proof that a designated
update was applied and sustains a device's identity across
updates of the recovery \gls{TCB}.

We demonstrate the effectiveness of \name on low-end devices by
implementing a prototype on an existing \gls{COTS} low-end
\gls{MCU} based on the new ARMv8-M processor family.
We make use of the isolation capabilities of TrustZone-M to
isolate the recovery \gls{TCB} from untrusted software and to
build an \gls{AWDT} without requiring additional hardware.
We make the following contributions:
\begin{itemize}
\item We design \name, a resilient \gls{IoT} architecture for
off-the-shelf \gls{IoT} devices. The design of \name respects
even the weakest class of \gls{IoT} devices and protects against
attackers actively aiming to render devices useless. With
\nametrigger, \name includes a zero-cost reset trigger, the
first of its kind. Our design relies on a \gls{TEE} which is
provided by modern processors.
\item We design \diceup which enables
updates of the recovery \gls{TCB} without loss of device
identity.
\item We implement a prototype on \name on a recent ARM Cortex-M33 based
\gls{MCU}, the \nxp \cite{LPC55S6x}. We leverage the new
TrustZone-M extension for ARM Cortex-M CPUs and ensure
portability across different ARMv8-M devices. Our implementation includes a full
setup featuring an IoT \hub, device provisioning and update
functionality.
\item We demonstrate how to protect an embedded OS, FreeRTOS \cite{freertos},
with our prototype. The OS modifications are limited to adding a
single supplementary task for extending \nametrigger in
coordination with the \hub.
\item We provide a security discussion and an evaluation of the
runtime and boot time performance impact as well as of the
memory requirements, showing the practicability of \name.
\end{itemize}

\section{Background}

This section provides background for the rest of the paper.

\subsection{Latches}

The code and data of \name on storage requires protection while
untrusted software executes.
Latches~\cite{Xu2019} can be used to protect critical code and
data from being overwritten or read out by untrusted software.
Conceptually, a latch is a state machine with two states, {\em
open} and {\em locked}.
A reset puts the latch into the open state. Software can put the
latch into the locked state (e.g., by writing to a hardware
register). Importantly,
the {\em only} action or event that can return the latch to the
open state is a reset. A latch has an associated
security function which allows some action (e.g., writing to
certain flash regions) if and only if the latch is in the open
state.

Latches allow trusted boot software that runs directly after a
reset to have full access to all hardware resources
but to selectively disable access to some of these resources to
(less trusted) software running subsequently on the device.
A read-write latch (RWLatch) can be used to protect secrets, for
instance, and a write latch (WRLatch) to protect the integrity
of data.

\subsection{DICE}

\gls{DICE} \cite{DICEoverview, DICEhwReq} is an industry
standard designed to enable attestation on low-end devices with
only minimal hardware support.
\gls{DICE} has been adopted by major \gls{MCU} manufacturers
such as NXP~\cite{lpc55} or Microchip~\cite{cec1702} and cloud
providers~\cite{AzureDice}.
\gls{DICE} is significantly more light-weight than alternatives
such as the TPM~\cite{tc2011tpm}. A DICE device has a unique
secret, the \gls{UDS} that is protected
by a latch. After a reset, DICE measures the first mutable
software component $\mathcal{M}$ (e.g., a boot loader) and uses
a one-way function to derive a symmetric key, called \gls{CDI},
from this measurement and the UDS. Next, \gls{DICE} makes the UDS inaccessible
until the next reset, discloses the \gls{CDI} to $\mathcal{M}$
and invokes $\mathcal{M}$.
This provides $\mathcal{M}$ with an identity that is unique to
both $\mathcal{M}$ and the device and that forms the foundation
for \gls{DICE} attestation.

Critically, the \gls{DICE} standard prescribes that the
computation of the \gls{CDI} must not be performed by mutable
software, i.e., by software that can be modified
or updated. While a device manufacturer may choose to implement
the computation of the \gls{CDI} in software, this software is
fixed (e.g., burned into ROM)
and not under the control of the device owner. This fact
together with the method by which the \gls{CDI} is derived
implies that (a) any modification of $\mathcal{M}$ will result
in a different \gls{CDI} and (b) the new \gls{CDI} is
unpredictable, as the \gls{UDS} is only known to the device's
DICE component, and the \gls{CDI} is derived from the \gls{UDS} via
a one-way function.

\gls{DICE} can be leveraged for lightweight device attestation,
for instance for \gls{IoT} devices connecting to a \hub
\cite{DICEattestation}.
$\mathcal{M}$ (or code $\mathcal{M}$ trusts) uses the \gls{CDI}
to derive two asymmetric key pairs, called the DeviceID and the
AliasID,
and uses the DeviceID private key to sign the AliasID
certificate.
Then, $\mathcal{M}$ removes all remnants of the \gls{CDI} and the
DeviceID private key and provides untrusted software only with
the AliasID key pair and the certificates.
This results in a cryptographic chain from the \gls{UDS} and
the \gls{CDI} to the DeviceID and the AliasID based on the state of a
device's software stack.
Untrusted software can use the AliasID as a device identifier for
authentication and attestation.

\subsection{\gls{MCU} Model}

Microcontroller units typically combine one or two low-end
microprocessor cores
with moderate amounts of RAM and flash memory and various simple
devices such
as watchdog timers and security features such as \gls{DICE}.
With the introduction
of ARMv8-M and TrustZone for Cortex-M, \glspl{TEE} are rapidly
becoming prevalent
even in the low-end \gls{MCU} market. 

This paper is based on an \gls{MCU} model whose processor(s)
support two privilege
levels (privileged and unprivileged) which are comparable to
user and kernel mode and which allow a simple ``operating system
kernel'' to protect itself from applications. In addition, our
\gls{MCU} model supports \gls{DICE} and, more importantly, a
\gls{TEE}. The assumed \gls{DICE} support is not critical
because,
as we will show, \name can implement \gls{DICE} in software. 

Our model of a \gls{TEE} is based on TrustZone for Cortex-M, the
{\em de facto} standard \gls{TEE} for \glspl{MCU}. Our
\gls{TEE} model isolates between two operating environments: a
higher-privileged trusted world and a lower-privileged normal
world. After a reset, boot code can assign various resources to
either the trusted or the normal world. This includes devices as
well as RAM and flash which can be assigned to either the
trusted or the normal world in chunks of different granularities. Code
running in the normal world can only access those resources that
have been assigned to it.

\section{Threat Model}
\label{sec:threat_model}

The software on the device is composed of \name and the main
\gls{IoT} application logic. The latter may include an OS and
applications running on it, a hypervisor, or bare-metal
applications and is subject to compromise by a \textit{remote}
attacker.
We assume the attacker to be capable of arbitrarily compromising
this untrusted software.
This includes the possibility that attacker code persists across
device resets.
An example is the exploitation of software vulnerabilities or
misconfigurations of the OS or application logic.
With these capabilities, the attacker is able to control common
\gls{IoT} devices at will and to make them unavailable.

\name is composed of its core TCB and a downloader, i.e. a
networking stack that allows \name to communicate directly with
the \hub. We assume the core TCB to be immune to compromise as
it is small and well isolated from a potential attacker. In
contrast, we only assume the more exposed downloader to work
correctly when not under attack. The core TCB puts protections
in place that allow easy recovery from a compromise of the
downloader.

We assume the hardware to work correctly according to
specification.
For instance, the attacker cannot alter code in
ROM, or interfere with peripherals shielded by the
\gls{TEE}.
However, the attacker may be able to leak secrets from the
\gls{TEE}, e.g., as demonstrated by Lapid and Wool \cite{Lapid2018}.
The attacker is unable to efficiently break state-of-the-art
cryptographic primitives.

The attacker may eavesdrop on the communication channel between
device and \hub as well as forge or tamper with messages.
However, the attacker can block the channel only for a limited
amount of time.
We consider prolonged attacks on the communication channel to be
detectable and remediatable by network operators.

The focus of this paper is the protection of \gls{IoT} devices.
The \hub, a web service likely to reside in a commercial cloud,
can be protected with the full array of industrial-strength IT
security solutions and is outside scope of this paper. Thus, we
consider the \hub to be immune to attacks. The same holds for
the environment in which devices are provisioned. As we focus on
scalable attacks, we leave physical attacks requiring proximity
of the attacker to a device out of scope.

\section{Design of \name}
\label{sec:concept_basic}

In this section, we describe the design of \name, for which we
combine several hardware and software building blocks.
To overcome the three problems of existing cyber-resilient
\gls{IoT} architectures listed in the introduction, we set the
following design goals:
\begin{description}
\item[DG-I\label{goal:r1}] Strong isolation of \name from
untrusted software while minimizing hardware requirements. This
makes \name applicable on a wide range of low-cost \gls{MCU}-based
\gls{IoT} devices.
\item[DG-II\label{goal:r2}] Prevent malware from damaging
devices or making them permanently unavailable.
\item[DG-III\label{goal:r3}] Provide a stable device identity
and attestation of the software stack even in case of an update
to \name.
\item[DG-IV\label{goal:r4}] \name should require only minimal
changes in existing OS and application software.
\end{description}

In support of DG-I, we will realize the reset trigger
\nametrigger and latches in software.
We achieve DG-II by regulating access by untrusted software to
peripherals critical for the availability of devices such as a
\gls{FMC}. \name requires only the following hardware:
\begin{description}
\item[Entropy source] to ensure freshness and non-forgeability
of messages \name exchanges with the \hub, e.g., a \gls{HRNG}.
\item[Orderly reset] to allow \name to execute deterministically
from a clean state after reset, regardless of prior state.
\item[\gls{TEE}] for realizing latches, \nametrigger, and for
controlling access to critical peripherals.
\item[Ordinary watchdog timer (WDT)] as a building block for
\nametrigger. This simple device is commonly included in
\glspl{MCU} to detect software crashes and reset the device.
Software is expected to service the WDT periodically. The WDT
will reset the device if it has not been serviced for some
amount of time.
\end{description}
These hardware features exist on a broad set of modern
\glspl{MCU}.
In contrast to CIDER \cite{Xu2019}, \name requires neither
hardware latches, nor an external reset trigger.

\name is split between a trusted boot loader and a runtime
component that executes inside the \gls{TEE} concurrently with
the untrusted software which runs outside the \gls{TEE}. After a
reset, the boot loader runs \diceup. Subsequently, similar to
the CIDER boot loader, it may download and install updates to
the untrusted software. Finally, it initializes the \name runtime
component in the \gls{TEE} and transfers control to the
untrusted software.

We first describe how we construct latches and \nametrigger to
isolate \name from untrusted software (\ref{goal:r1}).
Then, we focus on the isolation of critical peripherals
(\ref{goal:r2}) and on our extension of \gls{DICE}
(\ref{goal:r3}).
Finally, we describe how we use these mechanisms to design our
end-to-end system and refer to \ref{goal:r4}.

\subsection{Construction of Latches}

\name has to protect its binaries and data in flash from being
corrupted by the untrusted software. We configure the TEE to
make the flash range that stores \name code or data inaccessible
to the untrusted world. Effectively, this constitutes a latch, as
the restrictions will stay in place until the next reset, and
the untrusted software has no means of unlocking the latch. We
use the same mechanism to protect RAM for the \name runtime
component running in in the TEE.

\paragraph{DICE without hardware support:} On \glspl{MCU}
without hardware support for \gls{DICE}, we can use our
TEE-based latches to build \gls{DICE} in software. The software
consists of the \gls{UDS} (i.e., a unique secret) and code that
hashes the next binary and derives the \gls{CDI} from it and the
\gls{UDS}. This code runs immediately after a reset. Its last
action is to read-write-latch itself including the \gls{UDS},
thus making itself and the \gls{UDS} inaccessible until the next
reset. Control and the
\gls{CDI} are then transferred to the \name boot code.

\subsection{Construction of \nametrigger}

We use the \gls{TEE} and the WDT to construct our reset
trigger \nametrigger.
\nametrigger is software running in the secure world which
exposes the standard \gls{AWDT} interfaces \cite{Xu2019} to the
normal world.

A simple \nametrigger version could implement the \gls{AWDT}
initialization call by storing the \hub's public key (provided as a
parameter) and starting the hardware WDT with the timeout
value from the \gls{AWDT} initialization call. A second
\gls{AWDT} function generates a nonce using the \gls{HRNG},
stores it in trusted memory and also returns the nonce to the
untrusted caller. Untrusted application code can use the nonce
to request a {\em deferral ticket} for the \gls{AWDT} from the
\hub. If the \hub issues such a ticket, the untrusted
application code can use the third \gls{AWDT} interface to
request deferral of \gls{AWDT} expiration and, thus, device reset to be
postponed. \nametrigger could implement this third call by
using the public key that was provided during initialization to
verify the signature on the deferral ticket and only servicing
the WDT if the signature verification succeeds and the
nonce from the deferral ticket matches the stored nonce.

In practice, the situation is complicated by a mismatch between
the timeout intervals supported by most existing hardware
WDTs (at most few minutes) and those expected of
\glspl{AWDT} (hours to weeks). In order to support \gls{AWDT}
timeout intervals that are not limited by those of the
WDT hardware, we take advantage of a WDT feature
that causes the WDT to issue an interrupt a short time
before resetting the device. This interrupt warns software that
a device reset is imminent. We maintain a counter which is set
to a positive value during \gls{AWDT} initialization. Our
interrupt handler will decrement the counter and service the
WDT if and only if the result is not negative. For
example, if the WDT hardware only supports timeout
intervals of up to one minute, and we desire a 10 minute timeout
for the \gls{AWDT}, we can set the counter to 10 during
initialization or after a valid deferral ticket has been
received. This will cause \nametrigger to service the hardware
WDT ten times for a ten minute timeout interval.

\subsection{Isolation of Critical Peripherals}

\begin{figure}[t]
    \centering
    \includegraphics[width=\columnwidth]{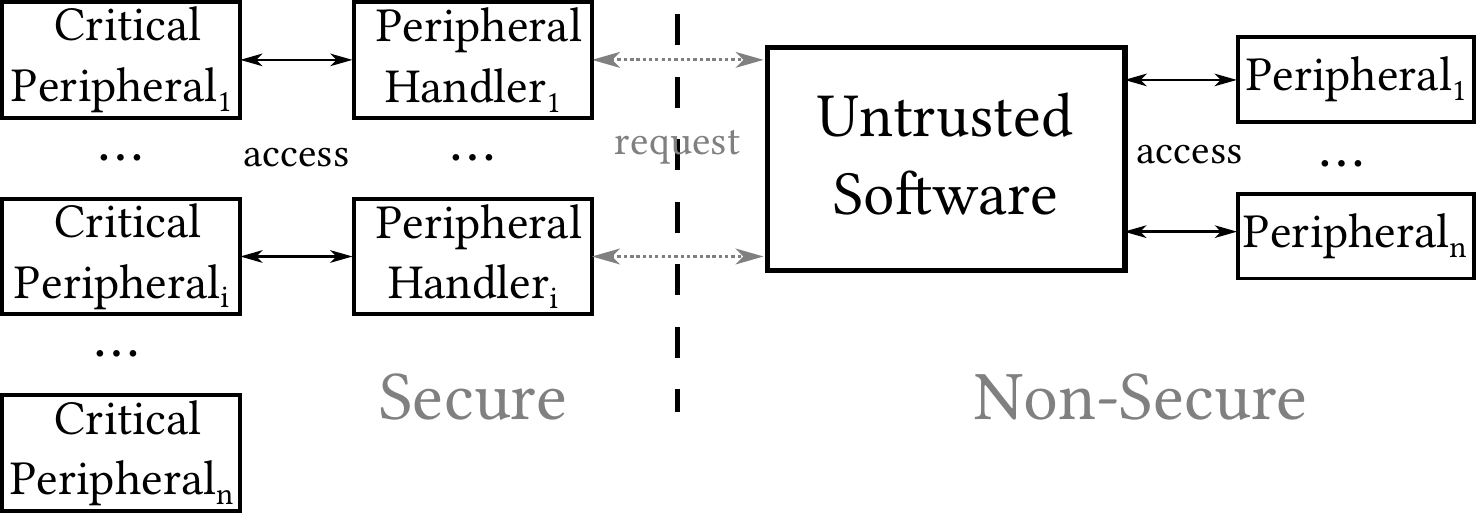}
    \caption{Interposition between untrusted software and critical peripherals through handlers in the \gls{TEE}.}
    \label{fig:concept-nscs}
\end{figure}

Untrusted software should be able to execute unconstrained and
thus be allowed to access peripherals.
However, to assure recoverability, we must regulate its access
to certain critical peripherals.
Otherwise, untrusted code could, for instance, put devices into
an irrecoverable low-power state, wear out flash, or permanently
disable peripherals used for communication.
This would make devices permanently unavailable and could even
cause physical damage.
Therefore, the \name runtime component running in the trusted
world of the TEE interposes between the untrusted code and these
critical peripherals.
We define a set of entry points into the \gls{TEE} to allow
untrusted software controlled access to critical peripherals.
\autoref{fig:concept-nscs} depicts this with ``peripheral
handlers'' serving as entry points into the \gls{TEE}.

An example of illicit use of critical peripherals is excessive
flash writes by untrusted software.
Even though the storage of trusted components of \name may be
latched using the \gls{TEE}, excessive writes to unprotected
storage locations could still cause flash to wear out.
flash writes are handled via an \gls{FMC}.
This is why we define the \gls{FMC} as a critical peripheral to
be only accessible from within the \gls{TEE}.
For persisting data, untrusted software must then use our
entry point to the \gls{TEE}, where our trusted \gls{FMC}
handler manages write requests.
Note that writing to flash is generally a slow operation.
Interposing on such operations will have a limited performance
impact.

Untrusted software might receive data from the
\hub, such as updates, or retrieve other data (e.g., from
sensors) it has to store.
When our flash write handler in the \gls{TEE} receives a flash
write attempt, it only allows:
\begin{itemize}
\item Writing to unprotected flash memory areas (areas that are
not latched). This includes areas storing code and data of
untrusted software and a staging area for storing updates.
Writes to latched areas (storing \name) are prohibited.
This
prevents untrusted software from circumventing the \gls{TEE}'s
WRLatch for \name code by using the handler.
\item A reasonable amount of flash writes. We assume that for
flash write events, a reasonable threshold or rate limitation to
regulate flash writes can be determined at design time of an
\gls{IoT} use case.
\end{itemize}

The \gls{FMC} is one example of a critical peripheral and its
presence or criticality may depend on the \gls{IoT} use case and
device type.
The absolute minimum set of critical peripherals that need to be
protected for \name are:
\begin{itemize}
\item Storage controllers allowing to write persistent storage
subject to wear out.
\item Power control peripherals that can put the device into  low power
modes or turn the device off.
 \item The WDT used by \nametrigger.
\end{itemize}
If a handler in the \gls{TEE} detects illicit use of
peripherals, it will reset the device.
This allows remediation through \name and can help the \hub
detect faulty behavior or compromise of business logic.
Other peripherals such as networking devices
or other storage controllers may be critical depending on the
device type and \gls{IoT} use case.
Our prototype demonstrates that adapting the untrusted software
to this protection mechanism requires only minor adaptations to
certain library function calls (\ref{goal:r4}).

\subsection{Extension of DICE}
\label{sec:concept_update}

To achieve design goal \ref{goal:r3}, we introduce a mechanism
that allows updating early-boot code like \name without loss of
device identity.
Updating early boot code like \name is essential in practice,
e.g. when updating cryptographic implementations or when an
exploitable vulnerability has been identified.
However, updating an early-boot component like \name results in
an unpredictable \gls{CDI} and thus in a new DeviceID and
AliasID.
A \hub knowing devices and their identities has no means of
relating the new and old DeviceID after an update and has no way
to verify that the update was indeed applied.

\begin{figure}[t]
    \centering
    \includegraphics[width=0.75\columnwidth]{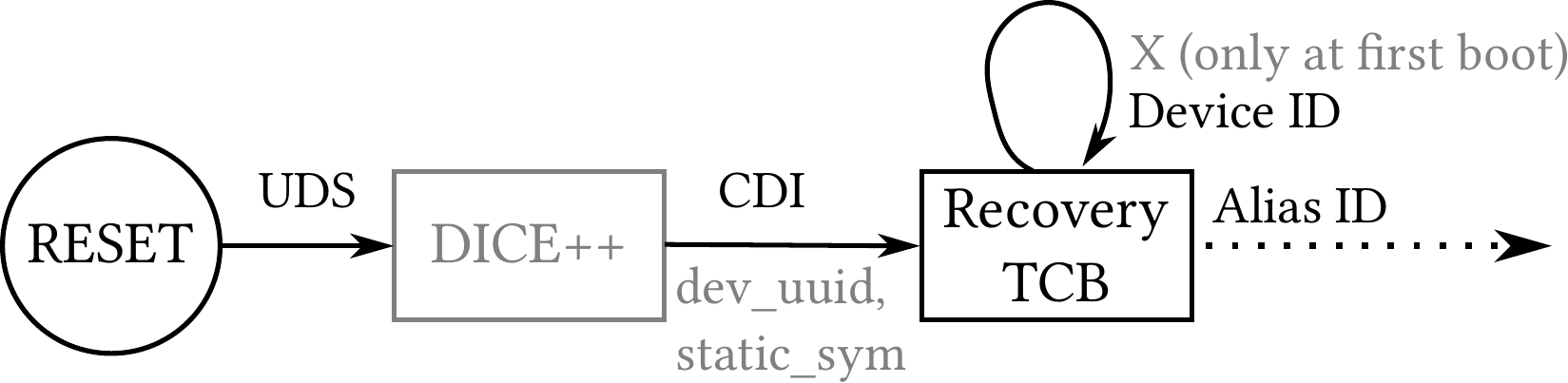}
    \caption{Modification to DICE (in gray) to exchange a shared secret and static identifier with the \hub at first device boot.}
    \label{fig:update-initial}
\end{figure}

As a practical solution for updating early-boot code without
loosing the device identity, we propose \diceup as an extension
to \gls{DICE}.
\diceup enables devices to provide the \hub with cryptographic
proof of identity after a change in the DeviceID.
This allows the \hub to associate the old and the new DeviceID
and AliasID and to verify that devices have properly applied
their early-boot code update.
\diceup thus minimizes the amount of code that cannot be updated in practice.
Our design of \diceup can supplement existing \gls{DICE}
implementations on shipped devices, which we demonstrate in
\autoref{sec:impl}.
The first part of this section describes device provisioning
while the second part explains device authentication after an
update of early-boot code.

\paragraph{Device Provisioning}

Our idea for re-associating device identities after an update of an early-boot component is to use a static secret and identifier shared between device and \hub.
We create the secret and identifier on the device in a secure environment during provisioning and exchange it with the \hub.
After an update, the device with a new DeviceID uses the static identifier for making an ``identity claim''.
The \hub requests proof of possession of the respective shared secret to verify a device's identity claim.

At the first boot of a device, \diceup randomly generates the static identifier $dev\_uuid$.
Further \diceup derives the shared secret $static\_sym$ as follows: $static\_sym := KDF(UDS, dev\_uuid)$.
The secret $static\_sym$ is derived with a one way function based on the \gls{UDS} and the static identifier, allowing no conclusions on the \gls{UDS}.
Both $dev\_uuid$ and $static\_sym$ remain unchanged when an early-boot component is updated.
We use a secret different than the \gls{UDS} to share with the \hub, because $static\_sym$ is less security critical.
$static\_sym$ is only used for re-association, not for regular attestation/authentication.
Device identities are not compromised should $static\_sym$ ever leak. In contrast, if the burnt-in \gls{UDS} leaks, the device is irrecoverably compromised.
In such cases, we can regenerate $dev\_UUID$, which causes derivation of a new $static\_sym$, avoiding loss of a device.

Being in a secure environment, $dev\_uuid$, $static\_sym$ and the initial DeviceID are read out and transferred to the \hub.
For this purpose, \diceup provides $dev\_uuid$ and $static\_sym$ once at first boot to the next layer.
At this time, early-boot code and environment can still be trusted.
It is also possible to exchange $dev\_uuid$ and $static\_sym$ online after enrollment.
In this case, the early-boot component generates the structure $X$ at initial boot and sends it to the \hub on first connection:
\[X:=sig(enc(dev\_uuid|static\_sym)^{Hub_{pub}})^{{DeviceID^1_{priv}}}\]
This means that the early-boot component encrypts $static\_sym$ and $dev\_uuid$ with the \hub public key and signs it with its initial DeviceID private key, enabling confidential and authenticated transmission.
The derivations at first boot are depicted in \autoref{fig:update-initial}.

\paragraph{Re-Association after Update}

\begin{figure}[t]
    \centering
    \includegraphics[width=\columnwidth]{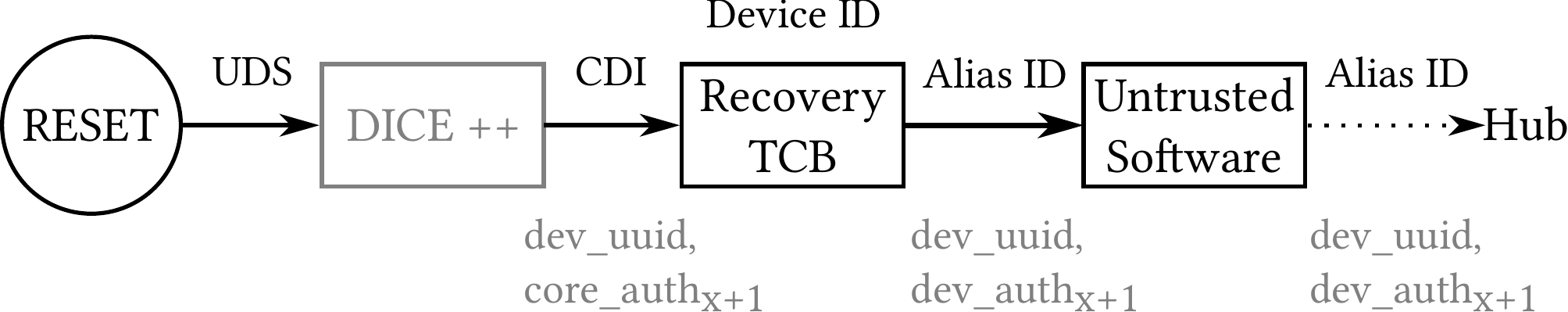}
    \caption{Modification to DICE key derivation and identity-based attestation (in gray) for updating the recovery TCB.}
    \label{fig:update-hub}
\end{figure}

At every boot, \diceup generates an additional key other than the \gls{CDI} for the early-boot component.
We call this key $core\_auth$, which \diceup derives as follows:
\[core\_auth_{x}:=KDF(static\_sym, hash(M_{x}|dev\_uuid))\]
This key depends on the current version $x$ of the early-boot component $\mathcal{M}$ on the device and allows it to authenticate itself.
\diceup hashes the early-boot component appended with the identifier $dev\_uuid$ and uses it with $static\_sym$ as input to a \gls{KDF}.
The key $core\_auth_{x}$ can only be computed with knowledge of $static\_sym$ and indicates that the software stack $\mathcal{M}$ in version $x$ was measured on the device with $dev\_uuid$.
An update to version $x+1$ causes an independent derivation of the key $core\_auth_{x+1}$.
Only \diceup and the \hub are able to derive this key as it depends on $static\_sym$ and thus cannot be forged or predicted.

The early-boot component receives this key from \diceup, which it uses to derive an ``identity token'', which we call $dev\_auth$.
Using $dev\_auth_{x+1}$, a modified early-boot component can prove to the \hub that it was successfully updated and is running on exactly the device with the prior DeviceID.
The derivations for $core\_auth$ and $dev\_auth$ are illustrated by \autoref{fig:update-hub}.
To compute $dev\_auth_{x+1}$, the early-boot component uses $core\_auth_{x+1}$ as key for a \gls{HMAC} over the new DeviceID and $dev\_uuid$, i.e.,
\[dev\_auth_{x+1}:=HMAC(core\_auth_{x+1}, DeviceID^{x+1}_{pub}|dev\_uuid)\]
$dev\_auth_{x+1}$ can only be computed by an early-boot component exactly in version $x+1$ running on the device $dev\_uuid$ and belongs to the DeviceID in version $x+1$.
$dev\_auth_{x+1}$ depends on $core\_auth_{x+1}$ which is only present after a correct update to version $x+1$.
Thus, $dev\_auth_{x+1}$ cannot be forged either.
Early-boot code can now pass $dev\_auth_{x+1}$ and $dev\_uuid$ to the next layer to handle the identity claim, for instance by integrating it into the DeviceID certificate.
When the \hub receives a claim of a device with DeviceID in version $x+1$ to be identified as device with $dev\_uuid$, the \hub requests the identity token.
The \hub computes the expected identity token on its side as all parameters are known to the \hub.
In case the received and expected token match, re-association is complete and the new DeviceID accepted.

\subsection{End-to-End System}

\begin{figure}[t]
    \centering
\includegraphics[width=\columnwidth]{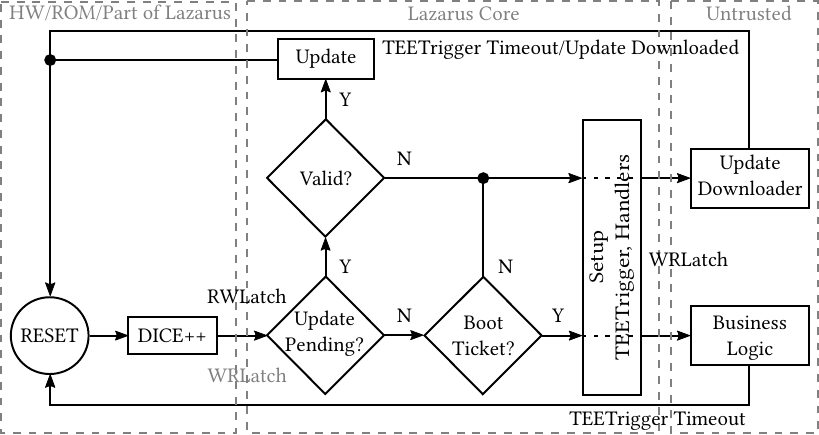}\caption{Boot flow of \name including latching, \nametrigger
initialization and peripheral handler setup.}
    \label{fig:backg-cider}
\end{figure}

This part describes how we compose our \gls{TEE}-based latching, our
\nametrigger, the protection mechanism for critical peripherals
and \diceup in our end-to-end system, \name.
We use CIDER \cite{Xu2019} as a basis for this.
\autoref{fig:backg-cider} illustrates the boot flow of \name.
Our first step is to apply temporal isolation, i.e., to execute
the core functionalities of \name \emph{before} untrusted
software like update download functionality or business logic
executes.
This means that the core part of \name executes directly after
an orderly reset.

After a reset, we execute \diceup (including \gls{DICE}, run by
the \gls{MCU}).
After activating the latches, \name executes its main
functionality ``\textit{\name Core}'' on the next lower security
tier in the \gls{TEE}.

\name Core checks whether an update for business logic or for
\name itself is pending, i.e., whether an update candidate was
stored in the ``staging area'' on persistent storage.
If so, \name Core verifies the integrity and authenticity of the
update using a public key it received during device provisioning
and a version number.
If the update is valid, \name Core applies the updates and
resets.
Otherwise, e.g., if it has been tampered with in the
communication channel, \name Core executes its ``update
downloader''.
The downloader contacts the \hub to retrieve authenticated
software.

The downloader, technically part of \name, has a significant attack surface, as it has to handle arbitrary packets from the internet.
\name Core thus considers it untrusted software.
This is why \name executes the update downloader in the same security tier as business logic outside the \gls{TEE}.
Before executing untrusted software, \name Core WRLatches itself including the downloader on storage.
This gives the update downloader the same storage protection as \name Core.
In addition, \name Core WRLatches the main memory region occupied by trusted runtime components running alongside untrusted software.
Further, \name Core initializes \nametrigger and constrains access to critical peripherals.
\name Core also derives the respective DeviceID and AliasID key material according to the \gls{DICE} specification for untrusted software.

In case no update is pending, \name Core checks for the presence of a ``boot ticket'' in the staging area.
The boot ticket is an accreditation of the \hub to boot one time into business logic without requiring further interaction with the \hub through the update downloader.
Without a valid boot ticket present, \name Core tries to acquire a boot ticket using the update downloader.
The downloader attests the device's software stack to the \hub by presenting the DeviceID and AliasID.
If the \hub is satisfied, it issues a boot ticket.
Like a deferral ticket, the boot ticket is a \hub-signed data structure that contains a nonce for ensuring freshness.
\name Core generates and stores a fresh nonce at each reset.
Upon reset, \name Core can verify the signature of a staged boot ticket and compare the contained nonce with the stored, old nonce.

If a valid boot ticket is present, \name Core boots into the business logic.
This stack may execute as long as it can acquire fresh deferral tickets from the \hub.
Business logic can also authenticate itself to the \hub using its AliasID and DeviceID credentials and may retrieve boot tickets.
Ticket acquisition from the \hub can be implemented as a task in an OS.

The only parts of \name active while untrusted software runs are \nametrigger and the logic regulating access to critical peripherals.
This code has no access to secrets, making side channel attacks pointless.
The \gls{TEE} exposes a small set of well-defined interfaces to allow untrusted software to interact with \nametrigger and to use critical peripherals.

Similar to \cite{Xu2019}, detection of compromise and identification of vulnerabilities is not in our scope.
We consider such lines of work orthogonal to our approach \cite{Williams2017}.
Availability of devices due to attacks on the communication channel or by making the \hub unavailable are discussed in \cite{Xu2019}.

\section{Implementation}
\label{sec:impl}

Our goal is to demonstrate that \name can be implemented even on
low-cost devices.
Therefore, we implemented a prototype for a low-end \gls{COTS}
\gls{MCU} from the \nxp series \cite{LPC55S6x}.
The price for \glspl{MCU} in this category ranges from cents to
a few dollars \cite{MCUMarket}.
While first \glspl{MCU} based on the ARMv8-M specification have
only recently been released, we expect this generation of
\glspl{MCU} to find wide adoption on the market.

We chose the LPCXpresso55S69 development
board~\cite{lpc55S69evk} as a target for our implementation. The
board is equipped with an NXP LPC55S69 (revision 0A) \gls{MCU}, peripherals
such as an accelerometer and several expansion ports. The
LPC55S69 \gls{MCU} features a dual core 32-bit ARM Cortex-M33
processor based on the ARMv8-M architecture \cite{armv8-m}
running at 96~MHz with TrustZone-M, 320~KB of SRAM, 640~KB of
flash, a watchdog timer, a \gls{HRNG}, crypto acceleration and
DICE support~\cite{LPC55S6x-um, nxpdice}.
Since our board does not include networking hardware, we attached an
off-the-shelf Wi-Fi chip, the ESP8266 \cite{ESP8266-ds}, to one of the
board's USARTs via one of the expansion ports. The ESP8266's serial
port connection to the USART runs at 115,200 baud, thus limiting the
network bandwidth to about 14 KB per second. 

With TrustZone-M, the Cortex-M CPU can be in secure or
non-secure mode. Each mode has a privileged and an unprivileged
level.
Non-secure execution can invoke the secure world through
\gls{NSC} functions.
Peripherals can be configured as fully secure, fully non-secure
or split into secure and non-secure partitions.
When the CPU is in a secure state, it can access both the secure
and the non-secure world.
In contrast, the processor can access only the non-secure
resources when it is in a non-secure state.

The \gls{MCU} hosts a secure AMBA AHB5 controller
\cite{amba,LPC55S6x-um} which allows configuring the whole memory map
including flash, RAM and peripherals as secure/non-secure and
privileged/unprivileged.
The ARM AMBA AHB5 protocol introduces signaling for secure and
non-secure transactions and therefore extends the TrustZone
technology from the processor to the entire system. \name can
be implemented on any ARMv8-M microcontroller that allows
securing our critical peripherals, which holds for every ARMv8-M
microcontroller currently on the market.

\begin{figure}[t]
    \centering
    \includegraphics[width=\columnwidth]{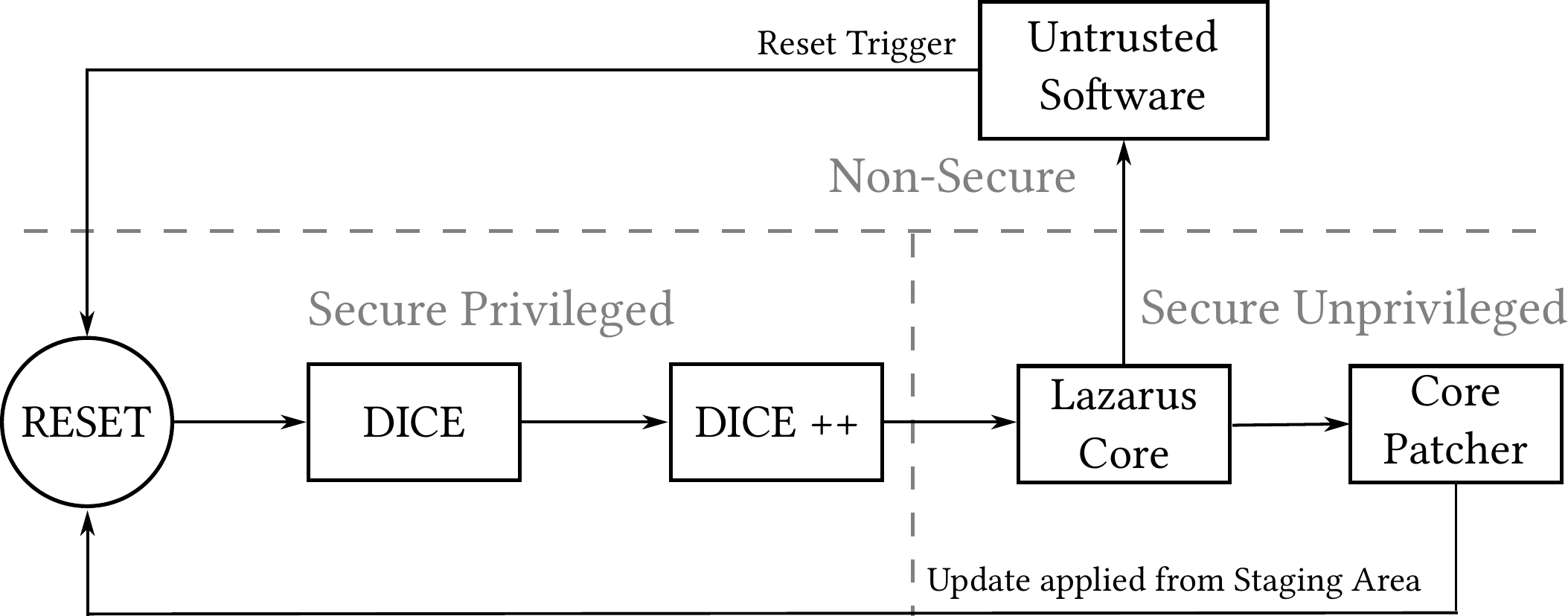}
    \caption{Overview of the implemented functionality along with execution modes and privilege levels.}
    \label{fig:update-overview}
\end{figure}

\begin{figure}[t]
    \centering
    \includegraphics[width=\columnwidth]{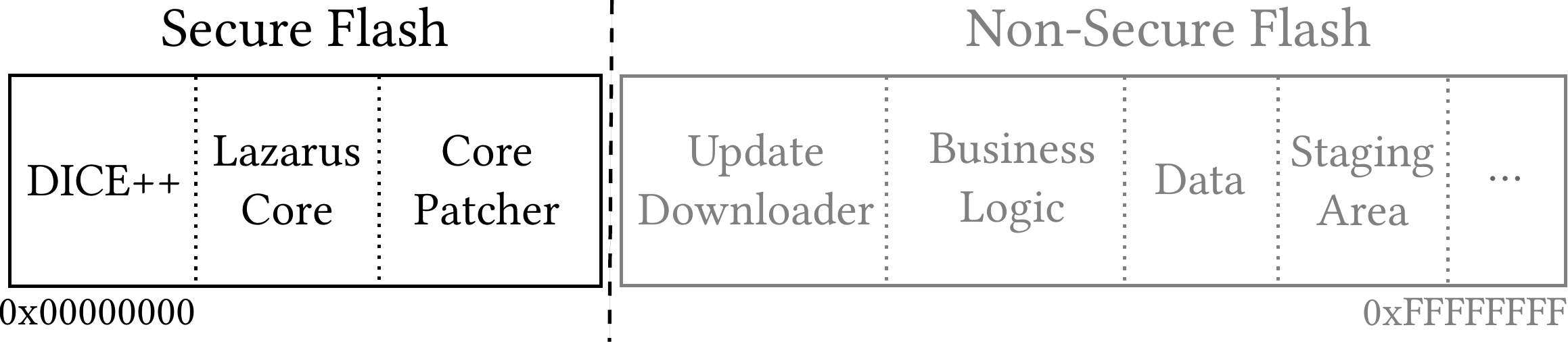}
    \caption{Layout of the binaries in flash memory along with their execution mode.}
    \label{fig:impl-flash}
\end{figure}

Our prototype splits the different functions of \name into
separate binaries running in different tiers as illustrated in
\autoref{fig:update-overview} and explained in the following.
The deployment of the binaries and their execution in secure and
non-secure mode led to the flash layout depicted in
\autoref{fig:impl-flash}.

We implemented the \name prototype in C. 
\autoref{tab:eval:loc} summarizes the \gls{LoC} of our
prototype, measured with Lizard \cite{lizard}.
The \name functionality executed in the secure world, 
consisting of \diceup, \namecore (\namecoretbl) and Core Patcher 
(\patchertbl), has less than 5k \gls{LoC} in total.
The update downloader (\udtbl) has about 1.4k \gls{LoC}.
In addition, our prototype uses parts of the RIoT crypto library
\cite{riotsource},
and parts of the CMSIS and NXP hardware-specific code.

Both CMSIS and NXP are relatively large bodies of headers and
support libraries
that implement a hardware abstraction layer for a large number
of different hardware
features of the \gls{MPU}. Our prototype uses only a small
number of these hardware
features and a similarly small fraction of the CMSIS and NXP
code. We excluded code
(e.g., header and C files) that our prototype obviously does not
use from the counts
in \autoref{tab:eval:loc}. Even so, the CMSIS and NXP numbers
still overestimate
the amount of code used in our prototype, as
separating all unused code would have been a complex task.

\begin{table}[t]
	\caption{LoC of the different components of our prototype.}
%    \small
	\centering
	\begin{tabular}{ccccccc}
		\toprule
		\dicepptbl & \namecoretbl & \patchertbl & \udtbl & Crypto & CMSIS & NXP \\
		\midrule
		1,668 & 2,753 & 571 & 1,359 & 3,564 &  1,764 & 5,682 \\
		\midrule
	\end{tabular}
	\label{tab:eval:loc}
\end{table}

\paragraph{DICE and DICE++}
As the \nxps supports \gls{DICE} according to the \gls{TCG}
specification \cite{DICEhwReq}, we do not have to implement
\gls{DICE} ourselves.
\gls{DICE} executes directly out of ROM after device reset
before any other software is executed.
For this reason, we cannot modify \gls{DICE}, which is why we
implemented \diceup as a separate binary executed right after
\gls{DICE}.
The \gls{MCU} executes the first code from flash after
\gls{DICE} from address \texttt{0x00000000}, which is where we
placed \diceup.
We calculate $static\_sym$ using HMAC-SHA256 at each boot.
Instead of using the \gls{UDS} (which we cannot access because
\gls{DICE} latches it), we derive $static\_sym$ from the
\gls{CDI} which \gls{DICE} passes to \diceup.
This does not weaken the security model because we remove the
\gls{CDI} before passing control to the next layer and hand over
only a re-derivation of the \gls{CDI}:
\[CDI':=HMAC(CDI, Hash(Lazarus\_Core))\]
\diceup derives $core\_auth$ and \gls{CDI}' using HMAC-SHA256.

For \diceup, we only need to protect $dev\_uuid$ and the \diceup
binary itself from being overwritten by the following layers.
To ensure this, we execute \diceup in secure privileged mode and
all following layers as either secure unprivileged or
non-secure.
We also use the secure AHB controller to WRLatch \diceup
including $dev\_UUID$.
The secure AHB controller can only be configured when executing
secure privileged.
Once activated, it cannot be re-configured or switched off until
the next reset.
This forces us to activate all required latches at once during
\gls{TEE} configuration.
For this reason, we configured the secure AHB controller in
\diceup to also WRLatch code and data of \name core that we
execute in secure unprivileged mode.
We also use the secure AHB controller to map the critical
peripherals into the secure world before executing untrusted
software.

\paragraph{\namecore}
\name Core is responsible for deriving the token
$dev\_auth$ for \name updates, for DeviceID and AliasID
derivation, for boot ticket verification, for boot nonce
generation, for checking if authorized code is present or
whether an update must be applied, for verifying and applying 
staged updates, and for regulating access to critical 
peripherals.

For secure communication with the \hub, \namecore derives the
DeviceID and AliasID from \gls{CDI}' and provides the signed
AliasID to untrusted software as input mapped to specific RAM
regions.
For these derivations, \name uses elliptic-curve cryptography
with the prime256v1 curve. The public keys are embedded
into X.509 certificates and the DeviceID certificate is signed
by the hub in order to establish a \gls{PKI}. For the
derivation of $dev\_auth$, \namecore uses HMAC-SHA256.

To regulate access to critical peripherals and to interact with
\nametrigger, \namecore sets up \gls{NSC} functions as an
interface for untrusted software.
We implemented several of these \gls{NSC} functions, i.e.,
{\em handlers}, for controlling access to critical peripherals.
For our implementation we considered the \gls{FMC}, the power
control peripheral, and the WDT as critical
\cite{LPC55S6x-um}.
The handler for the WDT implements the \nametrigger
deferral logic.
For deferral ticket nonce generation, the handler uses the
\gls{HRNG}.

Our handler for the power control peripheral ensures a device
can only be put into those low-power states that do not affect
\nametrigger.
Its underlying WDT is driven by the \gls{MCU}'s FRO\_1MHz clock.
Whether this clock is running depends on the power state of the
\gls{MCU}.
In active and sleep mode, it is enabled. However, in deep sleep,
the state of the clock can be configured off by software.
In the power-down modes the clock is off, which is why we block
deep sleep and power-down requests.

Our handler for the \gls{FMC} allows untrusted software only
writes to flash memory allocated to untrusted software and to
the staging area.
Untrusted software thus remains capable of persisting data to
specific flash regions, e.g., for providing boot and deferral
tickets or staging updates.
In addition, the handler implements a simple threshold scheme
that disallows excessive flash memory writes.
The protection of critical peripherals is not transparent to
untrusted software.
For this purpose, we modified the relevant libraries to use the
handler function instead of directly interacting with the
peripheral.

\paragraph{Core Patcher}
In general, \namecore is responsible for verifying and applying 
updates. However, as it cannot update itself in flash, in case a 
\namecore update is present on the staging area, \namecore only 
verifies this update and then invokes the Core Patcher. The Core 
Patcher's sole responsibility is to apply the \namecore update. 
Like \namecore, the Core Patcher runs in the secure world and can 
only be invoked by \namecore itself. This design makes it 
possible to update all binaries except \diceup.

\paragraph{Update Downloader}
The update downloader is part of \name, but as it implements a
networking stack for communicating with the \hub, it is
considered untrusted software.
This is why we execute it in the non-secure world.
Our handler for the \gls{FMC} peripheral ensures that the update
downloader can only write to specific areas in flash memory and
thus not overwrite itself or other parts of \name in case of
compromise.

We used the ESP8266 Wi-Fi chip \cite{ESP8266-ds} for TCP/IP
communication with the \hub and implemented our own format for
headers for binary updates.
We sign all messages exchanged between the device and the \hub
with their respective private keys and ensure the freshness of
critical messages using nonces for tickets and version numbers
for updates.
We implemented the update downloader as a bare-metal
application.
An advantage of our implementation is that the embedded OS \name
protects does not necessarily need to implement its own update
mechanism.

Our prototype treats the ESP8266 as a fixed-function peripheral
that is not subject to compromise. However, the ESP8266 has been
compromised in the past~\cite{10.1145/3375894.3375897}. The
ESP8266 is effectively another small microcontroller running its
own firmware stored in flash. It also exposes a firmware update
interface. We disabled this interface by connecting its GPIO0
pin to 3.3V.

Vulnerabilities in the ESP8266 firmware might still allow an
attacker to overwrite it and thus completely cut off our network
connection. Such attacks might originate with the untrusted
software or on the network. One can envision several defenses
based on \name or CIDER primitives to protect the ESP8266. For
example, a more robust version of the ESP8266 could write-latch
its firmware early in boot, such that it cannot be overwritten
even if a potential vulnerability is exploited. \name could also
support ESP8266 firmware updates and treat the ESP8266 as a
critical device by placing trusted handlers between the ESP8266
and untrusted software.

\paragraph{Business Logic}
We implemented two applications representing
real-world use cases. The first application is a traffic light
controller which controls LEDs via GPIO. The second application
periodically measures the temperature through an ADC sensor and
records the values in flash. Both applications run as tasks on
FreeRTOS.
We implemented the functionality for acquiring boot and deferral
tickets as a separate task in FreeRTOS.
This task uses the ESP8266 for the communication with the \hub.

\paragraph{Hub and Provisioning Environment}
Neither the \hub nor the provisioning environment are part of
the core \name device implementation.
However, in order to be able to perform a complete evaluation of
\name, we also implemented very simple
prototypes of a \hub and a provisioning environment.

We implemented the \hub in Python.
The \hub supports receiving AliasID certificates 
from the device, exchanging boot and deferral tickets and 
sending firmware updates. It is also able to issue an updated 
signed device configuration, e.g. to change Wi-Fi credentials, 
certificates or the \hub's address.

For device provisioning, we implemented a Python script that
builds (using \texttt{gcc-arm-none-eabi}) and flashes the
different binaries.
At first boot, the \gls{MCU} allows access to its memory via the
SWD Debug Interface using the NXP Link2 debugger to retrieve the
\gls{CSR} for the DeviceID certificate, $static\_sym$ and
$dev\_uuid$, and to deploy the signed DeviceID along with the
trust anchors for the \hub and for code signing.
The script also writes a device configuration data structure
containing initial Wi-Fi credentials and the \hub address.

\section{Evaluation}
\label{sec:eval}

In this section, we analyze our prototype regarding its code size, its boot and runtime performance impact, as well as regarding its networking overhead.
In the experiments in which the device communicates with the \hub, we ran our simple \hub prototype on a Lenovo Thinkpad T490s (with an Intel Core i7-8665U processor, 8~GB of DDR4 RAM and an Intel AC9560 wireless adapter) running Ubuntu 18.04 LTS and Python 3.6. The \hub and the device communicated via Wi-Fi through a lightly loaded AVM Fritz!Box 7362 SL wireless router.

\paragraph{Code size:} 
\autoref{tab:eval:sizes} shows the size of the \name binaries. 
\textit{Untrusted} is the part of \name that runs as part of the untrusted business 
logic, mainly the fetching of boot and deferral tickets.
We calculated the \textit{Untrusted} size as the difference between the 
sizes of the business logic binaries compiled with and without \name. 
In sum, \name takes up about 93~kB, which is about 14\% of the flash memory 
on the \nxps.
\name thus leaves enough resources for business logic.
Our prototype has not been subject to extensive code and binary optimization.
Not depicted in \autoref{tab:eval:sizes}, \name reserves 8~kB of RAM, i.e., 2.5\%, 
mapped to the secure world, unavailable to the non-secure world.

\begin{table}[t]
	\caption{Size in bytes of our binaries.}
	\centering
	\begin{tabular}{lccccc}
		\toprule
		Layer & \dicepptbl & \namecoretbl & \patchertbl & \udtbl & \textit{Untrusted} \\
		\midrule
		Size (bytes) & 10,720 & 43,156 & 7,676 & 22,332 & 1,640 \\
		\bottomrule
	\end{tabular}
	\label{tab:eval:sizes}
\end{table}

\paragraph{Boot time:}

We measured the boot time overhead as the time that elapses
from the execution of the first instruction of the \name boot
loader to the execution of the first instruction of the business
logic. We measured this time with an attached external device
as our measurement tool. We used a second \nxps \gls{MCU}.
We added instructions at the start and the end of the execution
of the boot loader to toggle a dedicated GPIO pin to signal the
measurement device to take a timestamp. We conducted 100
separate runs. We present average values only, as the standard
deviation was below 1\%.

The results are displayed in \autoref{tab:eval:boottime}. In the
case in which \namecore boots the business logic directly, we
measured a total overhead of less than 1.5 seconds. Most of this
time is spent computing cryptographic primitives, e.g., boot
ticket verification, in \namecore using the RIoT crypto library
\cite{riotsource} which does not use the \gls{MCU}'s crypto
accelerators. We consider this measurement a worst-case
estimate, showing that \name also performs well on \glspl{MCU}
without cryptographic accelerators. Drastic speed-ups can be
achieved using accelerators. For example, according to the
documentation for the \nxps, computing {\em ECDSA-secp256r1 Sign
+ Verify} takes only 116 ms \cite{LPC55S6x-um}.

Without a valid boot ticket, \namecore runs the update
downloader. In our scenario, the downloader contacts the \hub
with its AliasID, retrieves and stages a boot ticket and resets
the device. This takes around 3.4 seconds in total.

The third part of \autoref{tab:eval:boottime} displays the total
boot time if the \hub requires a firmware update to be
installed. We have omitted the individual stages of this
process, as large parts of it are a repetition of the second
case (i.e., no boot ticket). After booting through \diceup and
\namecore, the update downloader contacts the \hub and
receives instructions to download a firmware update. In the
experiment in \autoref{tab:eval:boottime}, it downloads and
stages a 58~KB update containing FreeRTOS and several
applications. After two more reset cycles, which install the
update and obtain a boot ticket, the device finally boots into
the newly installed firmware. As shown in
\autoref{tab:eval:boottime}, the entire process takes less than
16 seconds.
 
\begin{table}[t]
	\caption{Impact on boot time, depending on the presence of a valid boot ticket in the staging area.}
	\centering
	\begin{tabular}{lccc}
		\toprule
		Boot Path & Layer & Time Layer (ms) & Time Total (ms) \\
		\midrule
		w. Ticket & \dicepptbl & 152 & 152 \\
		& \namecoretbl & 1,297 & 1,449 \\
		\midrule
		w/o Ticket & \dicepptbl & 152 & 152 \\
		& \namecoretbl & 1,025 & 1,178 \\
		& \udtbl & 754 & 1,932 \\
		& \dicepptbl & 152 & 2,084 \\
		& \namecoretbl & 1,310 & 3,394 \\
		\midrule
		w. patch & &  & 15,720 \\
		\bottomrule
	\end{tabular}
	\label{tab:eval:boottime}
\end{table}

\paragraph{Runtime overhead:}
We measured the runtime overhead of \name against a
workload consisting of FreeRTOS running our two apps: the traffic light
controller and the ADC sensor app. For the baseline measurements,
we ran this workload unmodified and without the \name TrustZone
component interposing between the workload and the hardware.
For example, flash writes performed by the workload go directly to
flash memory.

The \name variant of the experiment runs the \name TrustZone component
which disables direct access by FreeRTOS and the applications to the flash
storage controller, the power controller and the watchdog timer.
As an alternative, \name exposes the TrustZone handler functions described in
\autoref{sec:impl}. The FreeRTOS workload is recompiled with libraries
that redirect accesses to the three devices to the TrustZone handlers. In addition,
we run a \nametrigger service application on FreeRTOS that obtains a deferral
ticket from the \hub over the network and passes the ticket to \nametrigger
running in TrustZone.

In each case, we used the FreeRTOS benchmarking feature to 
measure the CPU utilization. We compared the CPU idle time 
of the bare FreeRTOS app without \name to our \name implementation 
to obtain the total overhead.

Our measurements are summarized in \autoref{tab:eval:runtime}.
We measured a total overhead of 0.8\% when acquiring a deferral
ticket every minute with our task in FreeRTOS and verifying it
with \nametrigger.
A deferral ticket fetching interval of one or only a few minutes
is unreasonably short and was chosen only to demonstrate that
the overhead remains low even in this extreme case.
Fetching deferral tickets in intervals of multiple hours or days
is more realistic for most \gls{IoT} use cases and leads to only
negligible overhead.

In addition, we measured the overhead of only writing one flash
page (512~kB) with and without \name, i.e. writing to flash
via the handler versus writing to flash directly. This overhead
is about 0.19~\% and thus negligible.

 \begin{table}[t]
 \caption{Runtime performance impact for ticket fetching and verification with different deferral intervals.}
 \centering
  \begin{tabular}{lccccc}
     \toprule
		Ticket Fetch Interval & \SI{1}{\minute} & \SI{5}{\minute} & \SI{10}{\minute} & \SI{1}{\hour} & \SI{24}{\hour} \\
		\midrule
		Perf. Impact (\si{\percent}) & 0.80 & 0.16 & 0.08 & 0.01 & 0.001 \\
     \bottomrule
  \end{tabular}
 \label{tab:eval:runtime}
 \end{table}

In a second experiment, we measured the (wall clock) time
required for fetching and verifying a single deferral ticket.
We used the same measurement methodology as in the
boot time experiment, using an external device to take
time stamps based on GPIO signals. Once again, we repeated
each experiment 100 times.
\autoref{tab:eval:ticketingtime} shows the time spent on
computing tasks (Total) and the time that elapsed between
sending a ticket request to the \hub and receiving the ticket
(Network). The latter time does not affect the overhead, as
FreeRTOS can assign the processor to other tasks in the
meantime. The table also shows how the total compute time is
split between the TrustZone handler (Secure) and the untrusted
service application (Non-secure).

\begin{table}[t]
 \caption{Time required for ticket fetching and verification.}
 \centering
  \begin{tabular}{lcccc}
     \toprule
      & Total & Non-Secure & Secure & Network \\
      \midrule
		Fetch Time (ms) & 455 & 145 & 310 & 737 \\
     \bottomrule
  \end{tabular}
 \label{tab:eval:ticketingtime}
\end{table}

\paragraph{Networking:}
\autoref{tab:eval:ticketingsize} shows the TCP payload sizes for various 
messages exchanged with the \hub.
Requesting tickets is implemented as sending and receiving a \name 
authenticated header containing a type-field, nonce, digest and signature 
as well as a payload. In case of a boot ticket, the payload is just a fixed 
4 byte value, in case of a deferral ticket, the payload is the requested 
deferral time in milliseconds, also as a 4 byte value. 
This header is of size 140 bytes, totaling response and reply with payload 
to 288 bytes.
Sending the AliasID certificate to the \hub takes less than one kilobyte, 
depending on the actual size of the certificate, which can vary. 

The total TCP payload size of a boot in the update downloader path with 
subsequent ticket fetching in the business logic is less than 2,800 bytes, 
omitted from \autoref{tab:eval:ticketingsize}. 
This includes sending the AliasID and retrieving the boot ticket in the 
downloader, and subsequent sending of the AliasID and retrieval of both a 
deferral and boot ticket in the business logic.
After a normal boot with a valid boot ticket, data of less than 1,600 bytes 
are exchanged, i.e., sending the AliasID in the business logic and retrieving 
a boot and deferral ticket.
Subsequent fetches of deferral tickets require only 288 bytes.

\begin{table}[t]
	\caption{TCP payload sizes in bytes for different messages exchanged between device and \hub.}
	\centering
	\begin{tabular}{lccc}
		\toprule
		Message Type & Bytes Total & Request & Response \\
		\midrule
		Boot Ticket & 288 & 144 & 144 \\
		Deferral Ticket & 288 & 144 & 144 \\
		Send AliasID (max) & 1,003 & 1,001 & 2 \\
		\bottomrule
	\end{tabular}
	\label{tab:eval:ticketingsize}
\end{table}

\paragraph{Summary:}
To summarize, our evaluation shows that the overhead in terms of communication, flash and RAM requirements is modest even for resource-constrained devices.
The same holds for the runtime overhead for the communication with the \hub.
In contrast, the increase in boot times until executing business logic is not negligible.
However, we expect that resetting a device is an infrequent event, e.g., when a device needs to be serviced, or when the \hub suspects misbehavior.
The result is a good tradeoff between performance and security gain.
We seek to improve the required boot time with a more optimized implementation in future versions, e.g., using hardware cryptographic accelerators.

\section{Security Discussion}

Our goal is to ensure recoverability of \gls{IoT} devices even
if the untrusted software is completely compromised.
We achieve this goal if the \hub is able to enforce the software
stack the device executes within a time bound.
This means that soon after the \hub decides to deploy new
software, i.e., an update/patch, onto the device, our recovery
\gls{TCB} must be executed to retrieve and apply a new version.
We must assume that the attacker not only deploys malware that
tries to resist recovery but also tries to permanently render
devices useless.
Our attacker with full control over untrusted software and
possibly with access to the communication channel could try to
use the following vectors to attack availability:

\begin{description}
 \item[A-1\label{a:1}] Manipulate or block the communication channel between device and \hub.
 \item[A-2\label{a:2}] Attempt to tamper with \name by interfering with its execution, by overwriting it or by forging updates.
 \item[A-3\label{a:3}] Tamper with peripherals to permanently render devices irrecoverable. Examples are wear out of flash storage, shutting devices down or setting them into irrecoverable low-power states, or manipulating the firmware of other peripherals.
 \item[A-4\label{a:4}] Prevent or defer device reset, e.g, by manipulating or turning off \nametrigger or by forging deferral tickets.
 \item[A-5\label{a:5}] Inject malware into untrusted software and try to persist it across resets.
 \item[A-6\label{a:6}] Deceive the \hub about the application of updates to \name.
\end{description}

%a1
\ref{a:1} is only possible for a limited time. Our attack model
states that attacks on the communication channel can eventually
be detected and removed by the \gls{ISP}.
The effect of \gls{DoS} attacks on ticket fetching or \hub
availability were already discussed by Xu et al.~\cite{Xu2019} who observed
that \gls{DoS} attacks typically last less than 48 hours~\cite{istr}.

%a2, a3
To protect against \ref{a:2} and \ref{a:3}, we isolated the
trusted components and critical peripherals from untrusted
software.
We used a \gls{TEE} to latch trusted components and to map
critical peripherals into the secure world.
To enforce this, we simply configured TrustZone-M and the secure
AHB controller on our prototype.
Thus, an attacker is unable to misuse critical peripherals,
e.g., overwriting \name in flash, causing flash wear out,
deactivating \nametrigger's WDT, or by switching into an
irrecoverable low power state.

We kept the interface between the untrusted and the trusted components
very simple.
The runtime component contains only \nametrigger and the
handlers for critical peripherals.
\nametrigger only verifies a simple, fixed data structure, and
the flash write handler, merely evaluates whether flash writes
are excessive or target a certain area.
The runtime component does not process any secrets. Thus, our
system is by design not susceptible to side-channel attacks.
Untrusted input to \namecore processed during boot time must be
located in the staging area.
Every structure in the staging area has a well-defined format
and requires a valid signature from the \hub.
For the verification of all data structures, we provisioned the
device with public keys in a secure environment.
%

%a4
We prevent \ref{a:4} by placing \nametrigger inside the
\gls{TEE} and by using a simple interface into the \gls{TEE}.
This ensures the correct execution of \nametrigger which, by
construction, will force a device reset unless regularly
provided with a new deferral ticket from the \hub.
Deferral tickets are freshness-protected and only valid when
signed with the \hub private key.

%a5
Regarding \ref{a:5}, the attacker can compromise the untrusted
software according to our threat model.
The attacker's code can run on the device until the next reset.
The \hub can force such a reset
by refusing to issue \nametrigger deferral tickets (e.g.,
because it became aware of the attack).

If the attack was not persistent, the reset will remove it from
the device. The attacker may be able to
reinfect the device using the same method as the first time
(e.g., exploiting the same vulnerability).
Either way, the \hub can (and should) deploy a software update
that disables future attacks (e.g., by patching
the vulnerability). \name enables the \hub to do so irrespective
of the state of the device.

If the attack persisted itself, it must have changed storage.
If this change affected storage that is hashed by \gls{DICE}
(e.g., executable code), the hash values will also change.
Thus, attestation to the \hub will fail, and the \hub can force
a reset and update as in the nonpersistent case.

One can also envision data-only vulnerabilities that allow
an attacker to take control by changing data the hashed code
will read (e.g., configuration data read during OS boot)
 but that are not included in the hash because they can
change over time. 
While such persistent changes would not manifest themselves
in changed \gls{DICE} hashes, the device could still be recovered
unconditionally by using \name to patch the vulnerable code
and the exploit data.

If the downloader is free of flaws, it can retrieve updates
without obstacle.
However, we did not assume that the downloader is free of
vulnerabilities.
\name latches the downloader in flash, such that malware cannot
overwrite it.
It also activates \nametrigger before running the downloader.
This ensures that a reset will eventually occur.
A \gls{DoS} on the downloader is possible as long as it can be
re-compromised after each reset.
The \hub can detect such attacks (as it will fail to see device
attestations for the software update)
and block the attack with the help of the \gls{ISP}.

%a6
For defending against \ref{a:6}, we pointed out non-forgeability
and freshness of the relevant cryptographic messages sent
between the device and the \hub in \autoref{sec:concept_update}.
This only allows devices that indeed applied an update of a
trusted component to provide proof to the \hub.

In summary, \name allows a device to be reliably recovered and
its software (including \name) to be updated within a time bound
that can be configured by the \hub.

\section{Application in Real-World Use Cases}

Before applying \name in a real-world use case, the developer
has to configure \name based on the peripherals on the device
and based on the demands of the target application. This may
require writing new trusted handlers to regulate access to
critical peripherals and tailoring our prototype to concrete
\gls{IoT} scenarios, device types and specific peripherals,
e.g., sensors or additional flash storage attached via
\gls{SPI}.
In particular, all peripherals that can possibly be misused by
compromised business logic have to be guarded by \gls{TEE}
handlers.
For example, flash write handlers must implement a threshold or
rate limiting
to distinguish regular write activities from targeted flash wear
out.
In addition, business logic must be implemented such that it
does not exceed handler thresholds during regular operation.
This should be easy since the behavior of many \gls{IoT}
applications is quite regular and known in advance.

In addition to writing trusted handlers, the developer has to
modify the \gls{IoT} business logic
to call these handlers instead of trying to access the
peripherals directly. Such attempts would fail,
as the peripherals are protected by TrustZone from access by
untrusted software.

Further, integration of \name requires setting a reasonable
deferral timeout and determining the desired
behavior in case the device cannot contact the \hub, e.g.
because of a \gls{DoS} attack, as discussed in \cite{Xu2019}.
A possibility is to boot into a safe-mode version of the software with minimal
functionality as long as the \hub is unavailable.
However, this strongly depends on the \gls{IoT} use case the
system is tailored for.

\name can accommodate use cases that include code running in a
\gls{TEE} by running both the \name runtime component and the
business logic's \gls{TEE} code in the \gls{TEE}. Depending on
the size and the quality of the latter, this may increase the
attack surface of the \gls{TEE} code.

\section{Related Work}
\label{s:relwk}

Mechanisms for remote administration of servers have been
established in several industry standards~
\cite{ipmi,dcmi,redfish}.
These standards enable the efficient remote configuration,
monitoring and update of a large number of devices irrespective
of the state of their application processor and running software.
Devices are equipped with an additional co-processor able to
interrupt the application processor, and possibly with a
separate physical network interface.
This allows administrators to restore hung or compromised
systems, or even to activate servers that are shut down.
This achieves recoverability. However, the system design of adding
isolated co-processors and possibly separate network interfaces
is not an option for \gls{IoT} devices in the low-cost segment,
especially not for constrained devices.

Several mechanisms for automated malware detection and eviction
from end-user devices have been proposed
\cite{Nadji2011,Giffin2010, Hsu2006}.
This line of work focuses on security in a higher layer and has
a large \gls{TCB}, i.e., the OS kernel, a component \name
considers untrusted.
An architecture which leverages system virtualization for
automated detection and containment of rootkit attacks on Linux
systems was proposed by Baliga et al.~\cite{Baliga2006}.
\name also considers hypervisors untrusted components and can recover them.

%With Azure Sphere~\cite{hg2018azuresphere-overview,ms2018azuresphere-details} and 
Android Things~\cite{android-things} is a 
commercial secure \gls{IoT} platform.
It can be regarded as a full-fledged trustworthy \gls{IoT}
ecosystem with an OS and a bootloader for compliant device
architectures which enforces secure boot, runtime isolation,
attestation, hardware-based key protection, and handles remote
updates.
\gls{IoT} use cases are engineered on top of the platform and
can make use of the security services the OS provides, e.g., for
attestation or authentication.
The focus is on powerful Cortex-A devices.
There is, to the best of our knowledge, no means of recovering
from a compromised OS.
% or hypervisor.

For constrained devices, several secure update mechanisms have
been proposed, mainly for sensor networks in combination with
attestation~\cite{seshadri2004using,Seshadri:2006:SSC:1161289.1161306,
Perito:2010:SCU:1888881.1888931,7457156,sec-code-updates-mesh,cryptoeprint:2017:991,Ammar:2018:SSP:3176258.3176337}.
After trying to distribute the update, the server requests an
attestation proof of successful update application.
If the update fails (e.g., because malware is preventing it),
the compromised devices are revoked and require manual repair.
In contrast, \name permits remote recovery of all devices.
Some mechanisms allow wiping malware, e.g., with secure memory
erasure~\cite{Perito:2010:SCU:1888881.1888931,
Ammar:2018:SSP:3176258.3176337,7457156}.
However, common to all mechanisms is that they cannot reliably
recover from compromise of the host system and that they have
no provision to
force execution of recovery functionality.

Asokan et al.~\cite{2018arXiv180705002A} propose an architecture
for secure software updates on \glspl{MCU}.
They keep the secure update mechanism and key material in
read-only segments protected from the rest of the possibly
malicious system.
They built prototypes on two device types, HYDRA and Cortex-M.
The update functionality on HYDRA is loaded as the initial user
space process with highest priority after starting an seL4
kernel upon successful secure boot.
The isolation of the task's code and data is ensured by the seL4
kernel's separation.
In contrast, our goal is to decouple the recovery \gls{TCB} from
OSs or hypervisors.
The Cortex-M prototype uses TrustZone-M for isolation and
executes the update functionality as part of a trusted
bootloader.
Asokan et al. do not propose a mechanism that returns control to the
recovery \gls{TCB} after infection and 
provide no way to service the read-only update
functionality.

CIDER~\cite{Xu2019} is most closely related to our work and lays
the architectural
foundation for \name. CIDER appears to be targeted primarily at
higher-end devices for which security features such as storage
latches are common. The higher baseline cost of such devices
($\geq \$100$) also makes it reasonable to add missing security
support by attaching hardware costing a few dollars to an
extension port. CIDER did this for its external \gls{AWDT}. The
situation is fundamentally different for the cheap low-end
boards which are the focus of our work. While CIDER was also
implemented on a low-end MCU, this prototype inherits the
properties of its higher-end cousins, including an \gls{AWDT}
implementation on a separate MCU board. The paper mentions
cheaper alternatives without exploring them. In contrast, \name
demonstrates how a wide range of protections can be implemented
at zero cost in \gls{TEE} software. In addition, \name
provides protection against a range of attacks aiming to disable
the device and enables easy updates of the entire TCB.

%Cider, as already discussed, was implemented on high to low end boards, the SolidRun HummingBoard Edge \cite{HummingBoard}, Raspberry Pi 3 \cite{Rpi}, and STM Nucleo-L476RG \cite{STM32L476}.
%
%As motivated, Cider does not overcome major challenges, such as on recovery \gls{TCB} updates, protecting against malware trying to make devices unavailable.
%Further, CIDER imposes unrealistic requirements for resource-constrained devices and cannot be implemented on \gls{COTS} hardware.
%CIDER assumes the presence of storage latches.
%Further, CIDER uses an external chip as an \gls{AWDT}.
%That chip is an \gls{MCU} which has about the same complexity as small \gls{IoT} devices.
%
%Xu \etal mention in passing the possibility of using low-cost chips or a \gls{TEE} to build an \gls{AWDT}, but explore neither option.
%
The \gls{TCG} resilience work group provides no concrete implementation \cite{tcgcyres}. Auer et al. \cite{8714822} only mention integration of an \gls{AWDT} and recovery \gls{TCB} into a secure architecture for RISC-V.

%In summary, none of the architectures and mechanisms achieves our goals from \autoref{sec:concept_basic}.
%Most mechanisms fail to properly isolate the recovery \gls{TCB} (\ref{goal:r1}), e.g., when the \gls{TCB} is an OS or hypervisor and implement recovery as part of an OS or hypervisor (\ref{goal:r4}).
%Only CIDER implements a reset trigger at the cost of additional hardware requirements and makes assumption on the existence of latches, not generally available on devices (\ref{goal:r1}).
%None of the mechanisms explores adversaries trying to render devices useless (\ref{goal:r2}) or practical updates for the recovery \gls{TCB} itself (\ref{goal:r3}).

\section{Conclusion}
We present \name, a system for the recovery of compromised
low-end \gls{IoT} devices.
\name overcomes three major challenges in the design of
cyber-resilient \gls{IoT} architectures.
These challenges are applicability on low-cost \gls{COTS}
devices, defense against malware actively trying to make devices
unavailable, and practical updates of the recovery \gls{TCB}.
\name uses a \gls{TEE} to constrain malware.

As \glspl{TEE} are nowadays available even on low-cost devices,
\name can be deployed on a broad range of \gls{COTS} devices.
We use the \gls{TEE} to latch the recovery \gls{TCB}, isolate
our reset trigger \nametrigger and regulate access to critical
peripherals.
The latter prevents malware from rendering devices permanently
unavailable through misuse of peripherals, such as
entering irrecoverable sleep states or flash wear out, otherwise
possible within minutes~\cite{Templeman2012}.
For practical updates of the recovery \gls{TCB}, we propose an
extension to \gls{DICE}.
This extension allows sustaining the device identity through
secure re-association with the \hub.

We have implemented \name on a \gls{COTS} ARMv8-M \gls{MCU}
featuring TrustZone-M.
Our prototype has low memory requirements, negligible runtime
performance impact and modest boot time impact, making \name
suitable for adoption in a broad range of \gls{IoT} use cases.

%acm
\bibliographystyle{ACM-Reference-Format}
%llncs
%\balance
\bibliography{biblio}


\begin{thebibliography}{54}

%%% the Web address.

\ifx \showCODEN    \undefined \def \showCODEN     #1{\unskip}     \fi
\ifx \showDOI      \undefined \def \showDOI       #1{#1}\fi
\ifx \showISBNx    \undefined \def \showISBNx     #1{\unskip}     \fi
\ifx \showISBNxiii \undefined \def \showISBNxiii  #1{\unskip}     \fi
\ifx \showISSN     \undefined \def \showISSN      #1{\unskip}     \fi
\ifx \showLCCN     \undefined \def \showLCCN      #1{\unskip}     \fi
\ifx \shownote     \undefined \def \shownote      #1{#1}          \fi
\ifx \showarticletitle \undefined \def \showarticletitle #1{#1}   \fi
\ifx \showURL      \undefined \def \showURL       {\relax}        \fi
% invisible to TeX
\providecommand\bibfield[2]{#2}
\providecommand\bibinfo[2]{#2}
\providecommand\natexlab[1]{#1}
\providecommand\showeprint[2][]{arXiv:#2}

\bibitem[\protect\citeauthoryear{Ammar, Daniels, Crispo, and Hughes}{Ammar
  et~al\mbox{.}}{2018}]%
        {Ammar:2018:SSP:3176258.3176337}
\bibfield{author}{\bibinfo{person}{Mahmoud Ammar}, \bibinfo{person}{Wilfried
  Daniels}, \bibinfo{person}{Bruno Crispo}, {and} \bibinfo{person}{Danny
  Hughes}.} \bibinfo{year}{2018}\natexlab{}.
\newblock \showarticletitle{{SPEED: Secure Provable Erasure for Class-1 IoT
  Devices}}. In \bibinfo{booktitle}{\emph{Proceedings of the Eighth ACM
  Conference on Data and Application Security and Privacy}} (Tempe, AZ, USA)
  \emph{(\bibinfo{series}{CODASPY '18})}. \bibinfo{publisher}{ACM},
  \bibinfo{address}{New York, NY, USA}, \bibinfo{pages}{111--118}.
\newblock
\showISBNx{978-1-4503-5632-9}
\urldef\tempurl%
\url{https://doi.org/10.1145/3176258.3176337}
\showDOI{\tempurl}


\bibitem[\protect\citeauthoryear{{ARM}}{{ARM}}{2019}]%
        {amba}
\bibfield{author}{\bibinfo{person}{{ARM}}.} \bibinfo{year}{2019}\natexlab{}.
\newblock \bibinfo{title}{{AMBA 5 Overview}}.
\newblock
\newblock
\newblock
\shownote{\url{https://developer.arm.com/architectures/system-architectures/amba/amba-5}.}


\bibitem[\protect\citeauthoryear{{ARM Limited}}{{ARM Limited}}{2017}]%
        {armv8-m}
\bibfield{author}{\bibinfo{person}{{ARM Limited}}.}
  \bibinfo{year}{2017}\natexlab{}.
\newblock \bibinfo{title}{{Introduction to the ARMv8-M architecture}}.
\newblock
\newblock


\bibitem[\protect\citeauthoryear{{Asokan}, {Nyman}, {Rattanavipanon},
  {Sadeghi}, and {Tsudik}}{{Asokan} et~al\mbox{.}}{2018}]%
        {2018arXiv180705002A}
\bibfield{author}{\bibinfo{person}{N. {Asokan}}, \bibinfo{person}{T. {Nyman}},
  \bibinfo{person}{N. {Rattanavipanon}}, \bibinfo{person}{A.-R. {Sadeghi}},
  {and} \bibinfo{person}{G. {Tsudik}}.} \bibinfo{year}{2018}\natexlab{}.
\newblock \showarticletitle{{ASSURED: Architecture for Secure Software Update
  of Realistic Embedded Devices}}.
\newblock \bibinfo{journal}{\emph{ArXiv e-prints}} (\bibinfo{date}{July}
  \bibinfo{year}{2018}).
\newblock
\showeprint[arxiv]{1807.05002}~[cs.CR]


\bibitem[\protect\citeauthoryear{{Auer}, {Skubich}, and {Hiller}}{{Auer}
  et~al\mbox{.}}{2019}]%
        {8714822}
\bibfield{author}{\bibinfo{person}{L. {Auer}}, \bibinfo{person}{C. {Skubich}},
  {and} \bibinfo{person}{M. {Hiller}}.} \bibinfo{year}{2019}\natexlab{}.
\newblock \showarticletitle{A Security Architecture for RISC-V based IoT
  Devices}. In \bibinfo{booktitle}{\emph{2019 Design, Automation Test in Europe
  Conference Exhibition (DATE)}}. \bibinfo{pages}{1154--1159}.
\newblock
\showISSN{1530-1591}
\urldef\tempurl%
\url{https://doi.org/10.23919/DATE.2019.8714822}
\showDOI{\tempurl}


\bibitem[\protect\citeauthoryear{Baliga, Chen, and Iftode}{Baliga
  et~al\mbox{.}}{2006}]%
        {Baliga2006}
\bibfield{author}{\bibinfo{person}{Arati Baliga}, \bibinfo{person}{Xiaoxin
  Chen}, {and} \bibinfo{person}{Liviu Iftode}.}
  \bibinfo{year}{2006}\natexlab{}.
\newblock \showarticletitle{Paladin: Automated detection and containment of
  rootkit attacks}.
\newblock \bibinfo{journal}{\emph{Department of Computer Science, Rutgers
  University}} (\bibinfo{year}{2006}).
\newblock


\bibitem[\protect\citeauthoryear{Bogad and Huber}{Bogad and Huber}{2019}]%
        {10.1145/3375894.3375897}
\bibfield{author}{\bibinfo{person}{Katharina Bogad} {and}
  \bibinfo{person}{Manuel Huber}.} \bibinfo{year}{2019}\natexlab{}.
\newblock \showarticletitle{{Harzer Roller: Linker-Based Instrumentation for
  Enhanced Embedded Security Testing}}. In
  \bibinfo{booktitle}{\emph{Proceedings of the 3rd Reversing and
  Offensive-Oriented Trends Symposium}} (Vienna, Austria)
  \emph{(\bibinfo{series}{ROOTS’19})}. \bibinfo{publisher}{Association for
  Computing Machinery}, \bibinfo{address}{New York, NY, USA}, Article
  \bibinfo{articleno}{3}, \bibinfo{numpages}{9}~pages.
\newblock
\showISBNx{9781450377751}
\urldef\tempurl%
\url{https://doi.org/10.1145/3375894.3375897}
\showDOI{\tempurl}


\bibitem[\protect\citeauthoryear{Credencys}{Credencys}{2019}]%
        {vendingMachines}
\bibfield{author}{\bibinfo{person}{Credencys}.}
  \bibinfo{year}{2019}\natexlab{}.
\newblock \bibinfo{title}{{Increase your Vending Machine’s performance
  efficiency while bringing down the operational expenses \& maintenance cost
  with IoT}}.
\newblock
  \bibinfo{howpublished}{\url{https://www.credencys.com/smart-vending-machine-iot-solutions/}}.
\newblock


\bibitem[\protect\citeauthoryear{Daimler}{Daimler}{2019}]%
        {car2X}
\bibfield{author}{\bibinfo{person}{Daimler}.} \bibinfo{year}{2019}\natexlab{}.
\newblock \bibinfo{title}{{Networked with the surroundings. Car-to-X
  communication goes into series production}}.
\newblock
  \bibinfo{howpublished}{\url{https://www.daimler.com/innovation/case/connectivity/car-to-x-2.html}}.
\newblock


\bibitem[\protect\citeauthoryear{{Deogirikar} and {Vidhate}}{{Deogirikar} and
  {Vidhate}}{2017}]%
        {Deogirikar2017}
\bibfield{author}{\bibinfo{person}{J. {Deogirikar}} {and} \bibinfo{person}{A.
  {Vidhate}}.} \bibinfo{year}{2017}\natexlab{}.
\newblock \showarticletitle{Security attacks in IoT: A survey}. In
  \bibinfo{booktitle}{\emph{2017 International Conference on I-SMAC (IoT in
  Social, Mobile, Analytics and Cloud) (I-SMAC)}}. \bibinfo{pages}{32--37}.
\newblock
\showISSN{null}
\urldef\tempurl%
\url{https://doi.org/10.1109/I-SMAC.2017.8058363}
\showDOI{\tempurl}


\bibitem[\protect\citeauthoryear{{Distributed Management Task
  Force}}{{Distributed Management Task Force}}{2018}]%
        {redfish}
\bibfield{author}{\bibinfo{person}{{Distributed Management Task Force}}.}
  \bibinfo{year}{2018}\natexlab{}.
\newblock \bibinfo{title}{{Redfish Scalable Platforms Management API
  Specification v1.5}}.
\newblock
\newblock


\bibitem[\protect\citeauthoryear{DZone}{DZone}{2019}]%
        {smarthome}
\bibfield{author}{\bibinfo{person}{DZone}.} \bibinfo{year}{2019}\natexlab{}.
\newblock \bibinfo{title}{{Home Automation Using IoT}}.
\newblock
  \bibinfo{howpublished}{\url{https://dzone.com/articles/home-automation-using-iot}}.
\newblock


\bibitem[\protect\citeauthoryear{{Espressif Systems}}{{Espressif
  Systems}}{2019}]%
        {ESP8266-ds}
\bibfield{author}{\bibinfo{person}{{Espressif Systems}}.}
  \bibinfo{year}{2019}\natexlab{}.
\newblock \bibinfo{title}{{ESP8266EX Datasheet}}.
\newblock
\newblock


\bibitem[\protect\citeauthoryear{Feng, Qin, Zhao, Liu, Chu, and Feng}{Feng
  et~al\mbox{.}}{2017}]%
        {cryptoeprint:2017:991}
\bibfield{author}{\bibinfo{person}{Wei Feng}, \bibinfo{person}{Yu Qin},
  \bibinfo{person}{Shijun Zhao}, \bibinfo{person}{Ziwen Liu},
  \bibinfo{person}{Xiaobo Chu}, {and} \bibinfo{person}{Dengguo Feng}.}
  \bibinfo{year}{2017}\natexlab{}.
\newblock \bibinfo{title}{{Secure Code Updates for Smart Embedded Devices based
  on PUFs}}.
\newblock \bibinfo{howpublished}{Cryptology ePrint Archive, Report 2017/991}.
\newblock
\newblock
\shownote{\url{https://eprint.iacr.org/2017/991}.}


\bibitem[\protect\citeauthoryear{{Giffin}}{{Giffin}}{2010}]%
        {Giffin2010}
\bibfield{author}{\bibinfo{person}{J. {Giffin}}.}
  \bibinfo{year}{2010}\natexlab{}.
\newblock \showarticletitle{The Next Malware Battleground: Recovery After
  Unknown Infection}.
\newblock \bibinfo{journal}{\emph{IEEE Security Privacy}} \bibinfo{volume}{8},
  \bibinfo{number}{3} (\bibinfo{date}{May} \bibinfo{year}{2010}),
  \bibinfo{pages}{74--76}.
\newblock
\showISSN{1558-4046}
\urldef\tempurl%
\url{https://doi.org/10.1109/MSP.2010.107}
\showDOI{\tempurl}


\bibitem[\protect\citeauthoryear{Google}{Google}{2018}]%
        {android-things}
\bibfield{author}{\bibinfo{person}{Google}.} \bibinfo{year}{2018}\natexlab{}.
\newblock \bibinfo{title}{{Android Developers: Android Things}}.
\newblock
\newblock
\newblock
\shownote{\url{https://developer.android.com/things/}.}


\bibitem[\protect\citeauthoryear{HardwareBee}{HardwareBee}{2019}]%
        {MCUMarket}
\bibfield{author}{\bibinfo{person}{HardwareBee}.}
  \bibinfo{year}{2019}\natexlab{}.
\newblock \bibinfo{title}{{MCU Market History and Forecast 2016-2023}}.
\newblock
  \bibinfo{howpublished}{\url{http://hardwarebee.com/mcu-market-history-and-forecast-2016-2023/}}.
\newblock


\bibitem[\protect\citeauthoryear{Hilti}{Hilti}{2019}]%
        {hilti}
\bibfield{author}{\bibinfo{person}{Hilti}.} \bibinfo{year}{2019}\natexlab{}.
\newblock \bibinfo{title}{{White paper: Introducing Digital Asset Management}}.
\newblock
  \bibinfo{howpublished}{\url{https://www.hilti.com/content/dam/documents/pdf/w1/ontrack/whitepapers/W1_US_en_White\%20Paper\%20Increase\%20Profit\%20and\%20Productivity.pdf}}.
\newblock


\bibitem[\protect\citeauthoryear{{Hsu}, {Chen}, {Ristenpart}, {Li}, and
  {Su}}{{Hsu} et~al\mbox{.}}{2006}]%
        {Hsu2006}
\bibfield{author}{\bibinfo{person}{F. {Hsu}}, \bibinfo{person}{H. {Chen}},
  \bibinfo{person}{T. {Ristenpart}}, \bibinfo{person}{J. {Li}}, {and}
  \bibinfo{person}{Z. {Su}}.} \bibinfo{year}{2006}\natexlab{}.
\newblock \showarticletitle{Back to the Future: A Framework for Automatic
  Malware Removal and System Repair}. In \bibinfo{booktitle}{\emph{2006 22nd
  Annual Computer Security Applications Conference (ACSAC'06)}}.
  \bibinfo{pages}{257--268}.
\newblock
\showISSN{1063-9527}
\urldef\tempurl%
\url{https://doi.org/10.1109/ACSAC.2006.16}
\showDOI{\tempurl}


\bibitem[\protect\citeauthoryear{Huth, Duplys, and G{\"u}neysu}{Huth
  et~al\mbox{.}}{2016}]%
        {7457156}
\bibfield{author}{\bibinfo{person}{C. Huth}, \bibinfo{person}{P. Duplys}, {and}
  \bibinfo{person}{T. G{\"u}neysu}.} \bibinfo{year}{2016}\natexlab{}.
\newblock \showarticletitle{{Secure Software Update and IP Protection for
  Untrusted Devices in the Internet of Things via Physically Unclonable
  Functions}}. In \bibinfo{booktitle}{\emph{2016 IEEE International Conference
  on Pervasive Computing and Communication Workshops (PerCom Workshops)}}.
  \bibinfo{pages}{1--6}.
\newblock
\urldef\tempurl%
\url{https://doi.org/10.1109/PERCOMW.2016.7457156}
\showDOI{\tempurl}


\bibitem[\protect\citeauthoryear{{Intel Corporation}}{{Intel
  Corporation}}{2011}]%
        {dcmi}
\bibfield{author}{\bibinfo{person}{{Intel Corporation}}.}
  \bibinfo{year}{2011}\natexlab{}.
\newblock \bibinfo{title}{{Data Center Manageability Interface Specification
  v1.5 rev. 1.0}}.
\newblock
\newblock


\bibitem[\protect\citeauthoryear{{Intel Corporation}}{{Intel
  Corporation}}{2013}]%
        {ipmi}
\bibfield{author}{\bibinfo{person}{{Intel Corporation}}.}
  \bibinfo{year}{2013}\natexlab{}.
\newblock \bibinfo{title}{{Intelligent Platform Management Interface
  Specification v2.0 rev. 1.1}}.
\newblock
\newblock


\bibitem[\protect\citeauthoryear{Kohnh{\"a}user and
  Katzenbeisser}{Kohnh{\"a}user and Katzenbeisser}{2016}]%
        {sec-code-updates-mesh}
\bibfield{author}{\bibinfo{person}{Florian Kohnh{\"a}user} {and}
  \bibinfo{person}{Stefan Katzenbeisser}.} \bibinfo{year}{2016}\natexlab{}.
\newblock \showarticletitle{{Secure Code Updates for Mesh Networked Commodity
  Low-End Embedded Devices}}. In \bibinfo{booktitle}{\emph{Computer Security --
  ESORICS 2016}}, \bibfield{editor}{\bibinfo{person}{Ioannis Askoxylakis},
  \bibinfo{person}{Sotiris Ioannidis}, \bibinfo{person}{Sokratis Katsikas},
  {and} \bibinfo{person}{Catherine Meadows}} (Eds.).
  \bibinfo{publisher}{Springer International Publishing},
  \bibinfo{address}{Cham}, \bibinfo{pages}{320--338}.
\newblock
\showISBNx{978-3-319-45741-3}


\bibitem[\protect\citeauthoryear{Lapid and Wool}{Lapid and Wool}{2018}]%
        {Lapid2018}
\bibfield{author}{\bibinfo{person}{Ben Lapid} {and} \bibinfo{person}{Avishai
  Wool}.} \bibinfo{year}{2018}\natexlab{}.
\newblock \bibinfo{title}{Cache-Attacks on the ARM TrustZone implementations of
  AES-256 and AES-256-GCM via GPU-based analysis}.
\newblock \bibinfo{howpublished}{Cryptology ePrint Archive, Report 2018/621}.
\newblock
\newblock
\shownote{\url{https://eprint.iacr.org/2018/621}.}


\bibitem[\protect\citeauthoryear{Media}{Media}{2019}]%
        {sierra}
\bibfield{author}{\bibinfo{person}{OpenSystems Media}.}
  \bibinfo{year}{2019}\natexlab{}.
\newblock \bibinfo{title}{{IoT Based Smart Traffic Signal Monitoring Using
  Vehicle Count}}.
\newblock
  \bibinfo{howpublished}{\url{https://www.embedded-computing.com/guest-blogs/iot-based-smart-traffic-signal-monitoring-using-vehicle-count}}.
\newblock


\bibitem[\protect\citeauthoryear{{Microchip}}{{Microchip}}{2020}]%
        {cec1702}
\bibfield{author}{\bibinfo{person}{{Microchip}}.}
  \bibinfo{year}{2020}\natexlab{}.
\newblock \bibinfo{title}{{CEC1702}}.
\newblock
\newblock
\newblock
\shownote{\url{https://www.microchip.com/wwwproducts/en/CEC1702}.}


\bibitem[\protect\citeauthoryear{{Microsoft Research}}{{Microsoft
  Research}}{2019}]%
        {riotsource}
\bibfield{author}{\bibinfo{person}{{Microsoft Research}}.}
  \bibinfo{year}{2019}\natexlab{}.
\newblock \bibinfo{title}{{Robust Internet of Things}}.
\newblock \bibinfo{howpublished}{\url{https://github.com/microsoft/RIoT}}.
\newblock


\bibitem[\protect\citeauthoryear{Nadji, Giffin, and Traynor}{Nadji
  et~al\mbox{.}}{2011}]%
        {Nadji2011}
\bibfield{author}{\bibinfo{person}{Yacin Nadji}, \bibinfo{person}{Jonathon
  Giffin}, {and} \bibinfo{person}{Patrick Traynor}.}
  \bibinfo{year}{2011}\natexlab{}.
\newblock \showarticletitle{Automated Remote Repair for Mobile Malware}. In
  \bibinfo{booktitle}{\emph{Proceedings of the 27th Annual Computer Security
  Applications Conference}} (Orlando, Florida, USA)
  \emph{(\bibinfo{series}{ACSAC '11})}. \bibinfo{publisher}{ACM},
  \bibinfo{address}{New York, NY, USA}, \bibinfo{pages}{413--422}.
\newblock
\showISBNx{978-1-4503-0672-0}
\urldef\tempurl%
\url{https://doi.org/10.1145/2076732.2076791}
\showDOI{\tempurl}


\bibitem[\protect\citeauthoryear{{Nawir}, {Amir}, {Yaakob}, and {Lynn}}{{Nawir}
  et~al\mbox{.}}{2016}]%
        {Nawir2016}
\bibfield{author}{\bibinfo{person}{M. {Nawir}}, \bibinfo{person}{A. {Amir}},
  \bibinfo{person}{N. {Yaakob}}, {and} \bibinfo{person}{O.~B. {Lynn}}.}
  \bibinfo{year}{2016}\natexlab{}.
\newblock \showarticletitle{Internet of Things (IoT): Taxonomy of security
  attacks}. In \bibinfo{booktitle}{\emph{2016 3rd International Conference on
  Electronic Design (ICED)}}. \bibinfo{pages}{321--326}.
\newblock
\showISSN{null}
\urldef\tempurl%
\url{https://doi.org/10.1109/ICED.2016.7804660}
\showDOI{\tempurl}


\bibitem[\protect\citeauthoryear{NetScout}{NetScout}{2019}]%
        {mirai}
\bibfield{author}{\bibinfo{person}{NetScout}.} \bibinfo{year}{2019}\natexlab{}.
\newblock \bibinfo{title}{{Mirai IoT Botnet Description and DDoS Attack
  Mitigation}}.
\newblock
  \bibinfo{howpublished}{\url{https://www.netscout.com/blog/asert/mirai-iot-botnet-description-and-ddos-attack-mitigation}}.
\newblock


\bibitem[\protect\citeauthoryear{{NXP}}{{NXP}}{2019a}]%
        {nxpdice}
\bibfield{author}{\bibinfo{person}{{NXP}}.} \bibinfo{year}{2019}\natexlab{a}.
\newblock \bibinfo{title}{{AN12278 - LPC55S69 Security Solutions for IoT}}.
\newblock
\newblock


\bibitem[\protect\citeauthoryear{{NXP}}{{NXP}}{2019b}]%
        {lpc55S69evk}
\bibfield{author}{\bibinfo{person}{{NXP}}.} \bibinfo{year}{2019}\natexlab{b}.
\newblock \bibinfo{title}{{LPC55S69-EVK: LPCXpresso55S69 Development Board}}.
\newblock
\newblock
\newblock
\shownote{\url{https://www.nxp.com/products/processors-and-microcontrollers/arm-microcontrollers/general-purpose-mcus/lpc5500-cortex-m33/lpcxpresso55s69-development-board:LPC55S69-EVK}.}


\bibitem[\protect\citeauthoryear{NXP}{NXP}{2019}]%
        {LPC55S6x}
\bibfield{author}{\bibinfo{person}{NXP}.} \bibinfo{year}{2019}\natexlab{}.
\newblock \bibinfo{title}{{LPC55S6x: High Efficiency Arm® Cortex®-M33-based
  Microcontroller Family}}.
\newblock
  \bibinfo{howpublished}{\url{https://www.nxp.com/products/processors-and-microcontrollers/arm-microcontrollers/general-purpose-mcus/lpc5500-cortex-m33/high-efficiency-arm-cortex-m33-based-microcontroller-family:LPC55S6x}}.
\newblock


\bibitem[\protect\citeauthoryear{{NXP}}{{NXP}}{2019a}]%
        {lpc55}
\bibfield{author}{\bibinfo{person}{{NXP}}.} \bibinfo{year}{2019}\natexlab{a}.
\newblock \bibinfo{title}{{LPC55Sxx Secure Boot}}.
\newblock
\newblock
\newblock
\shownote{\url{https://www.nxp.com/docs/en/application-note/AN12283.pdf}.}


\bibitem[\protect\citeauthoryear{{NXP}}{{NXP}}{2019b}]%
        {LPC55S6x-um}
\bibfield{author}{\bibinfo{person}{{NXP}}.} \bibinfo{year}{2019}\natexlab{b}.
\newblock \bibinfo{title}{{UM11126 LPC55S6x/LPC55S2x/LPC552x User manual}}.
\newblock
\newblock


\bibitem[\protect\citeauthoryear{Perito and Tsudik}{Perito and Tsudik}{2010}]%
        {Perito:2010:SCU:1888881.1888931}
\bibfield{author}{\bibinfo{person}{Daniele Perito} {and} \bibinfo{person}{Gene
  Tsudik}.} \bibinfo{year}{2010}\natexlab{}.
\newblock \showarticletitle{{Secure Code Update for Embedded Devices via Proofs
  of Secure Erasure}}. In \bibinfo{booktitle}{\emph{Proceedings of the 15th
  European Conference on Research in Computer Security}} (Athens, Greece)
  \emph{(\bibinfo{series}{ESORICS'10})}. \bibinfo{publisher}{Springer-Verlag},
  \bibinfo{address}{Berlin, Heidelberg}, \bibinfo{pages}{643--662}.
\newblock
\showISBNx{3-642-15496-4, 978-3-642-15496-6}
\urldef\tempurl%
\url{http://dl.acm.org/citation.cfm?id=1888881.1888931}
\showURL{%
\tempurl}


\bibitem[\protect\citeauthoryear{Raghavendra, Sivalingam, and
  Znati}{Raghavendra et~al\mbox{.}}{2006}]%
        {raghavendra2006wireless}
\bibfield{author}{\bibinfo{person}{Cauligi~S Raghavendra},
  \bibinfo{person}{Krishna~M Sivalingam}, {and} \bibinfo{person}{Taieb Znati}.}
  \bibinfo{year}{2006}\natexlab{}.
\newblock \bibinfo{booktitle}{\emph{Wireless sensor networks}}.
\newblock \bibinfo{publisher}{Springer}.
\newblock


\bibitem[\protect\citeauthoryear{Sanuel}{Sanuel}{2017}]%
        {AzureDice}
\bibfield{author}{\bibinfo{person}{Arjmand Sanuel}.}
  \bibinfo{year}{2017}\natexlab{}.
\newblock \bibinfo{title}{Azure {IoT} supports new security hardware to
  strengthen {IoT} security}.
\newblock
\newblock
\newblock
\shownote{\url{https://azure.microsoft.com/en-us/blog/azure-iot-supports-new-security-hardware-to-strengthen-iot-security/}.}


\bibitem[\protect\citeauthoryear{Services}{Services}{2019}]%
        {freertos}
\bibfield{author}{\bibinfo{person}{Amazon~Web Services}.}
  \bibinfo{year}{2019}\natexlab{}.
\newblock \bibinfo{title}{{freeRTOS}}.
\newblock \bibinfo{howpublished}{\url{https://www.freertos.org/}}.
\newblock


\bibitem[\protect\citeauthoryear{Seshadri, Luk, Perrig, Doorn, and
  Khosla}{Seshadri et~al\mbox{.}}{2004}]%
        {seshadri2004using}
\bibfield{author}{\bibinfo{person}{Arvind Seshadri}, \bibinfo{person}{Mark
  Luk}, \bibinfo{person}{Adrian Perrig}, \bibinfo{person}{Leendert~van Doorn},
  {and} \bibinfo{person}{Pradeep Khosla}.} \bibinfo{year}{2004}\natexlab{}.
\newblock \bibinfo{title}{{Using FIRE \& ICE for Detecting and Recovering
  Compromised Nodes in Sensor Networks}}.
\newblock
\newblock


\bibitem[\protect\citeauthoryear{Seshadri, Luk, Perrig, van Doorn, and
  Khosla}{Seshadri et~al\mbox{.}}{2006}]%
        {Seshadri:2006:SSC:1161289.1161306}
\bibfield{author}{\bibinfo{person}{Arvind Seshadri}, \bibinfo{person}{Mark
  Luk}, \bibinfo{person}{Adrian Perrig}, \bibinfo{person}{Leendert van Doorn},
  {and} \bibinfo{person}{Pradeep Khosla}.} \bibinfo{year}{2006}\natexlab{}.
\newblock \showarticletitle{{SCUBA: Secure Code Update By Attestation in Sensor
  Networks}}. In \bibinfo{booktitle}{\emph{Proceedings of the 5th ACM Workshop
  on Wireless Security}} (Los Angeles, California) \emph{(\bibinfo{series}{WiSe
  '06})}. \bibinfo{publisher}{ACM}, \bibinfo{address}{New York, NY, USA},
  \bibinfo{pages}{85--94}.
\newblock
\showISBNx{1-59593-557-6}
\urldef\tempurl%
\url{https://doi.org/10.1145/1161289.1161306}
\showDOI{\tempurl}


\bibitem[\protect\citeauthoryear{Sigfox}{Sigfox}{2019}]%
        {smart_livestock}
\bibfield{author}{\bibinfo{person}{Sigfox}.} \bibinfo{year}{2019}\natexlab{}.
\newblock \bibinfo{title}{{Smart livestock collars let ranchers track, monitor
  and manage herds like never before}}.
\newblock
  \bibinfo{howpublished}{\url{https://www.sigfox.com/en/solutions/smart-livestock-collars-let-ranchers-track-monitor-and-manage-herds-never}}.
\newblock


\bibitem[\protect\citeauthoryear{{Symantec}}{{Symantec}}{2016}]%
        {istr}
\bibfield{author}{\bibinfo{person}{{Symantec}}.}
  \bibinfo{year}{2016}\natexlab{}.
\newblock \bibinfo{title}{{Internet Security Threat Report}}.
\newblock
\newblock


\bibitem[\protect\citeauthoryear{Symantec}{Symantec}{2019}]%
        {hajime}
\bibfield{author}{\bibinfo{person}{Symantec}.} \bibinfo{year}{2019}\natexlab{}.
\newblock \bibinfo{title}{{Hajime worm battles Mirai for control of the
  Internet of Things}}.
\newblock
  \bibinfo{howpublished}{\url{https://www.symantec.com/connect/blogs/hajime-worm-battles-mirai-control-internet-things}}.
\newblock


\bibitem[\protect\citeauthoryear{{TC Group and others}}{{TC Group and
  others}}{2011}]%
        {tc2011tpm}
\bibfield{author}{\bibinfo{person}{{TC Group and others}}.}
  \bibinfo{year}{2011}\natexlab{}.
\newblock \bibinfo{title}{{TPM Main Specification Version 1.2 Rev. 116}}.
\newblock
\newblock
\newblock
\shownote{\url{https://trustedcomputinggroup.org/resource/tpm-main-specification/}.}


\bibitem[\protect\citeauthoryear{{Templeman} and {Kapadi}}{{Templeman} and
  {Kapadi}}{2012}]%
        {Templeman2012}
\bibfield{author}{\bibinfo{person}{R. {Templeman}} {and} \bibinfo{person}{A.
  {Kapadi}}.} \bibinfo{year}{2012}\natexlab{}.
\newblock \showarticletitle{{GANGRENE}: Exploring the Mortality of Flash
  Memory}. In \bibinfo{booktitle}{\emph{Presented as part of the 7th {USENIX}
  Workshop on Hot Topics in Security}}. \bibinfo{publisher}{{USENIX}},
  \bibinfo{address}{Bellevue, WA}.
\newblock
\urldef\tempurl%
\url{https://www.usenix.org/conference/hotsec12/workshop-program/presentation/Templeman}
\showURL{%
\tempurl}


\bibitem[\protect\citeauthoryear{{Terry Yin}}{{Terry Yin}}{2019}]%
        {lizard}
\bibfield{author}{\bibinfo{person}{{Terry Yin}}.}
  \bibinfo{year}{2019}\natexlab{}.
\newblock \bibinfo{title}{{A simple code complexity analyzer without caring
  about the C/C++ header files or Java imports.}}
\newblock \bibinfo{howpublished}{\url{https://terryyin.github.io/lizard/}}.
\newblock


\bibitem[\protect\citeauthoryear{Tiwari, Ballal, and Lewis}{Tiwari
  et~al\mbox{.}}{2007}]%
        {Tiwari:2007:EWS:1210669.1210670}
\bibfield{author}{\bibinfo{person}{Ankit Tiwari}, \bibinfo{person}{Prasanna
  Ballal}, {and} \bibinfo{person}{Frank~L. Lewis}.}
  \bibinfo{year}{2007}\natexlab{}.
\newblock \showarticletitle{Energy-efficient Wireless Sensor Network Design and
  Implementation for Condition-based Maintenance}.
\newblock \bibinfo{journal}{\emph{ACM Trans. Sen. Netw.}} \bibinfo{volume}{3},
  \bibinfo{number}{1}, Article \bibinfo{articleno}{1} (\bibinfo{date}{March}
  \bibinfo{year}{2007}).
\newblock
\showISSN{1550-4859}
\urldef\tempurl%
\url{https://doi.org/10.1145/1210669.1210670}
\showDOI{\tempurl}


\bibitem[\protect\citeauthoryear{{Trusted Computing Group}}{{Trusted Computing
  Group}}{2017}]%
        {DICEoverview}
\bibfield{author}{\bibinfo{person}{{Trusted Computing Group}}.}
  \bibinfo{year}{2017}\natexlab{}.
\newblock \bibinfo{title}{{Foundational Trust for IoT and Resource Constrained
  Devices}}.
\newblock
\newblock


\bibitem[\protect\citeauthoryear{{Trusted Computing Group}}{{Trusted Computing
  Group}}{2018a}]%
        {DICEhwReq}
\bibfield{author}{\bibinfo{person}{{Trusted Computing Group}}.}
  \bibinfo{year}{2018}\natexlab{a}.
\newblock \bibinfo{title}{{Hardware Requirements for a Device Identifier
  Composition Engine}}.
\newblock
\newblock


\bibitem[\protect\citeauthoryear{{Trusted Computing Group}}{{Trusted Computing
  Group}}{2018b}]%
        {DICEattestation}
\bibfield{author}{\bibinfo{person}{{Trusted Computing Group}}.}
  \bibinfo{year}{2018}\natexlab{b}.
\newblock \bibinfo{title}{{Implicit Identity Based Device Attestation}}.
\newblock
\newblock


\bibitem[\protect\citeauthoryear{{Trusted Computing Group}}{{Trusted Computing
  Group}}{2019}]%
        {tcgcyres}
\bibfield{author}{\bibinfo{person}{{Trusted Computing Group}}.}
  \bibinfo{year}{2019}\natexlab{}.
\newblock \bibinfo{title}{{TCG Cyber Resilient Technologies}}.
\newblock
  \bibinfo{howpublished}{\url{https://trustedcomputinggroup.org/wp-content/uploads/TCG-Cyber-Resilient-Technologies-\%E2\%80\%93-Rob-Spiger-Microsoft.pdf}}.
\newblock


\bibitem[\protect\citeauthoryear{{Williams}, {McMahon}, {Samtani}, {Patton},
  and {Chen}}{{Williams} et~al\mbox{.}}{2017}]%
        {Williams2017}
\bibfield{author}{\bibinfo{person}{R. {Williams}}, \bibinfo{person}{E.
  {McMahon}}, \bibinfo{person}{S. {Samtani}}, \bibinfo{person}{M. {Patton}},
  {and} \bibinfo{person}{H. {Chen}}.} \bibinfo{year}{2017}\natexlab{}.
\newblock \showarticletitle{Identifying vulnerabilities of consumer Internet of
  Things (IoT) devices: A scalable approach}. In \bibinfo{booktitle}{\emph{2017
  IEEE International Conference on Intelligence and Security Informatics
  (ISI)}}. \bibinfo{pages}{179--181}.
\newblock
\showISSN{null}
\urldef\tempurl%
\url{https://doi.org/10.1109/ISI.2017.8004904}
\showDOI{\tempurl}


\bibitem[\protect\citeauthoryear{Xu, Huber, Sun, England, Peinado, Lee,
  Marochko, Mattoon, Spiger, and Thom}{Xu et~al\mbox{.}}{2019}]%
        {Xu2019}
\bibfield{author}{\bibinfo{person}{Meng Xu}, \bibinfo{person}{Manuel Huber},
  \bibinfo{person}{Zhichuang Sun}, \bibinfo{person}{Paul England},
  \bibinfo{person}{Marcus Peinado}, \bibinfo{person}{Sangho Lee},
  \bibinfo{person}{Andrey Marochko}, \bibinfo{person}{Dennis Mattoon},
  \bibinfo{person}{Rob Spiger}, {and} \bibinfo{person}{Stefan Thom}.}
  \bibinfo{year}{2019}\natexlab{}.
\newblock \showarticletitle{{Dominance as a New Trusted Computing Primitive for
  the Internet of Things}}. In \bibinfo{booktitle}{\emph{Proceedings of the
  2019 IEEE Symposium on Security and Privacy}} (San Francisco, CA)
  \emph{(\bibinfo{series}{SP '19})}. \bibinfo{publisher}{IEEE Computer
  Society}, \bibinfo{address}{Washington, DC, USA}.
\newblock


\end{thebibliography}
 
\end{document}